\documentclass[english]{article}
\usepackage{lmodern}

\usepackage[T1]{fontenc}
\usepackage[latin9]{inputenc}
\usepackage{mathrsfs}
\usepackage{amsmath}
\usepackage{amssymb}
\usepackage{mathdots}
\usepackage{graphicx}
\usepackage{wasysym}
\usepackage{booktabs}

\makeatletter
\newcommand{\lyxaddress}[1]{
\par {\raggedright #1
\vspace{1.4em}
\noindent\par}
}

\usepackage{mathrsfs}   
\usepackage{slashed}     
\usepackage{bbold}  
\usepackage{dsfont}
\usepackage{url}
\usepackage{graphicx}
\usepackage[colorlinks=true,linkcolor=redLinks,citecolor=greenLinks,urlcolor=redLinks, pdfborder={0 0 1}]{hyperref}
\usepackage{color} 
\usepackage{xcolor} 
\usepackage{framed}

\usepackage[numbers,sort&compress]{natbib}

\definecolor{greenLinks}{rgb}{0, 0.6, 0} 
\definecolor{blueLinks}{rgb}{0, 0, 0.6}
\definecolor{redLinks}{rgb}{0.6, 0, 0}
\definecolor{eprintLinks}{rgb}{0.4, 0.4, 0.4}
\definecolor{tempText}{rgb}{0.55, 0.10,0.67}
\definecolor{journalLinks}{rgb}{0.6, 0, 0}

\colorlet{shadecolor}{gray!15}

\newcommand{\MYhref}[3][redLinks]{\href{#2}{\color{#1}{#3}}}%

\usepackage{multirow}
\let\orig@Hy@EveryPageAnchor\Hy@EveryPageAnchor
\def\Hy@EveryPageAnchor{%
    \begingroup
    \hypersetup{pdfview=Fit}%
    \orig@Hy@EveryPageAnchor
    \endgroup
}

\let\oldFootnote\footnote
\newcommand\nextToken\relax

\renewcommand\footnote[1]{%
    \oldFootnote{#1}\futurelet\nextToken\isFootnote}

\newcommand\isFootnote{%
    \ifx\footnote\nextToken\textsuperscript{,}\fi}

\makeatother

\usepackage{babel}
\begin{document}

\title{{\Large{}\vspace{-1.0cm}} \hfill {\normalsize{}IFIC/17-45} \\*[10mm]
  {\huge{}Quasi-Dirac neutrino oscillations}{\Large{}\vspace{0.5cm}}}

\author{{\Large{}Gaetana Anamiati}\thanks{E-mail: anamiati@ific.uv.es}
  {\Large{}, Renato M. Fonseca}\thanks{E-mail: renato.fonseca@ific.uv.es}
  {\Large{}, Martin Hirsch}\thanks{E-mail: mahirsch@ific.uv.es} \date{}}

\maketitle

\lyxaddress{\begin{center}
{\Large{}\vspace{-0.5cm}}AHEP Group, Instituto de Física Corpuscular,
C.S.I.C./Universitat de València\\
Edifício de Institutos de Paterna, Apartado 22085, E--46071 València,
Spain
\par\end{center}}

\begin{center}
\date
\par\end{center}
\begin{abstract}

Dirac neutrino masses require two distinct neutral
Weyl spinors per generation, with a special arrangement of masses and
interactions with charged leptons. Once this arrangement is perturbed,
lepton number is no longer conserved and neutrinos become Majorana
particles. If these lepton number violating perturbations are small
compared to the Dirac mass terms, neutrinos are {\em quasi-Dirac}
particles.  Alternatively, this scenario can be characterized by the
existence of pairs of neutrinos with almost degenerate masses, and a
lepton mixing matrix which has 12 angles and 12 phases.  In this
work we discuss the phenomenology of quasi-Dirac neutrino oscillations
and derive limits on the relevant parameter space from various
experiments.  In one parameter perturbations of the Dirac limit, very
stringent bounds can be derived on the mass splittings between the
almost degenerate pairs of neutrinos.  However, we also demonstrate
that with suitable changes to the lepton mixing matrix, limits on such
mass splittings are much weaker, or even completely absent. Finally,
we consider the possibility that the mass splittings are too small to
be measured and discuss bounds on the new, non-standard lepton mixing
angles from current experiments for this case.

\end{abstract}
\newpage{}

\section{Introduction}
\label{sect:Intro}

Neutrino oscillation experiments cannot distinguish Dirac from
Majorana neutrinos, hence it is still unknown whether or not lepton
number is conserved. Other processes, such as neutrinoless double beta
decay \cite{Avignone:2007fu,Deppisch:2012nb}, need to be probed in
order to answer this question. However, while the nature of neutrinos
is often seen as a dichotomy, presenting two sharply distinct
scenarios, the Dirac neutrino case can be seen as a limit of the
more general Majorana case in which lepton number violating mass terms
are zero, and this limit can be approached smoothly.

In practice, one can start with a model of $2n$ Majorana neutrinos and
get a phenomenology arbitrarily close to the one of a model of $n$
Dirac neutrinos. This can already be seen with only one
generation of active ($\nu$) and sterile neutrinos ($N^{c}$). In the
basis $\left(\nu,N^{c}\right)^{T}$ the most general mass matrix reads:
\begin{equation}\label{eq:MassMat}
	m_{\nu}=\left(\begin{array}{cc}
		m_{L} & m_{D}\\
		m_{D} & m_{R}
	\end{array}\right)\,.
\end{equation}
If $m_{L}=m_{R}=0$, lepton number is preserved and neutrinos are Dirac
particles. This limit can alternatively be characterized by two {\em
	exactly} degenerate mass eigenstates composed in equal parts of
$\nu$ and $N^c$: $\nu_{1}=1/\sqrt{2}\left(\nu+N^{c}\right)$ and
$\nu_{2}=i/\sqrt{2}\left(-\nu+N^{c}\right)$.\footnote{Note the factor
	$i$ in $\nu_{2}$. One could equally well choose the two mass
	eigenstates to be $\pm m_D$ instead.}  Small deviations from the
limit $m_{L}=m_{R}=0$ lead to a {\em quasi-Dirac} scenario where
lepton number is no longer exactly preserved.

Let us rewrite eq. (\ref{eq:MassMat}) using:
\begin{eqnarray}\label{eq:epsdel}
	\varepsilon=\frac{(m_{L}+m_{R})}{2 m_D},
	\\ 
	\theta=\frac{(m_{L}-m_{R})}{4 m_D}.
\end{eqnarray}
As long as $\varepsilon$ and $\theta$ are much smaller than one,
we obtain:
\begin{align}
	m_{1,2} & \simeq m_{D}\left(1\pm\varepsilon\right)\,,\label{eq:2}\\
	\nu_{1} & \simeq 1/\sqrt{2}
	\left[\left(1+\theta\right)\nu+\left(1-\theta\right)N^{c}\right]\,,\\
	\nu_{2} & \simeq i/\sqrt{2}
	\left[\left(-1+\theta\right)\nu+\left(1+\theta\right)N^{c}\right]\,.\label{eq:4}
\end{align}
Departures from the Dirac case therefore can manifest themselves as
either new mass splittings or new mixing angles (or, in general,
both). Moreover, as this simple example shows, mass splittings and
mixing angles are completely independent of each other.  Note that for
small values of $\varepsilon$ and $\theta$, lepton number violation is
naturally suppressed, as expected. This can be most easily seen in our
one generation scenario for the double beta decay observable $\langle
m_{\nu}\rangle$: for $\theta=0$ ($\varepsilon=0$) it is straightforwardly
calculated to be $\langle m_{\nu}\rangle\simeq \varepsilon m_D$ ($\langle
m_{\nu}\rangle \simeq 2 \theta m_D$).

We have therefore the following situation. Oscillation experiments
cannot distinguish a model with $n$ Majorana neutrinos (containing $n$ Weyl
spinors) from one with $n$ Dirac neutrinos (containing $2n$ Weyl spinors) with
matching masses and mixing angles. Nevertheless, once we add to a
model with Dirac neutrinos small sources of lepton number violation,
oscillation probabilities will change. Some illustrative examples are
shown in fig. (\ref{fig:PeeRe}).  We plot there the electron
neutrino survival probability for low-energy (reactor) neutrinos at
distances up to (and slightly larger than) the typical distances of
the KamLAND experiment \cite{Abe:2008aa}. In all plots the black lines
show the expectation for the current global best fit point
\cite{deSalas:2017kay} for the ordinary neutrino parameters in the
standard three generation case, to which we have added either a
non-zero mass splitting to a Dirac state (top row) or one particular
new quasi-Dirac angle (bottom row).

In section \eqref{sect:param} we will discuss the general
parametrization of masses and mixing angles for scenarios with three
generations of quasi-Dirac neutrinos. However, from the examples shown
in fig. (\ref{fig:PeeRe}) one can read off already some basic facts
about oscillations of quasi-Dirac neutrinos, which we will work out in
greater detail in section (\ref{sect:exp}). First, small non-zero
values of $\varepsilon$'s are equivalent to introducing new, large
oscillation lengths.  Thus, the best constraints on $\varepsilon$ will
come from oscillation experiments with the largest possible
baselines. And secondly, even if mass splittings are negligibly small,
the new, non-standard angles which appear in this setup (called
$\theta$ above) may affect oscillation probabilities in a way similar
to standard angles, hence creating parameter degeneracies. For
example, as fig. (\ref{fig:PeeRe}) shows, from $P_{ee}$ alone one
cannot provide limits on a single angle. (In this example variations
of $\theta_{14}$ can be compensated by varying $\theta_{12}$.) Even by
combining more than one oscillation probability, constraints can only
be derived for certain combinations of angles and phases of the mixing
matrix. We will discuss this in detail in section
(\ref{subsect:angs}). Constraints on mass splittings are discussed in
section (\ref{subsect:splits}).

\begin{center}
	\begin{figure}[tbph]
		\begin{centering}
			\hspace{0.5cm}\includegraphics[scale=0.52]{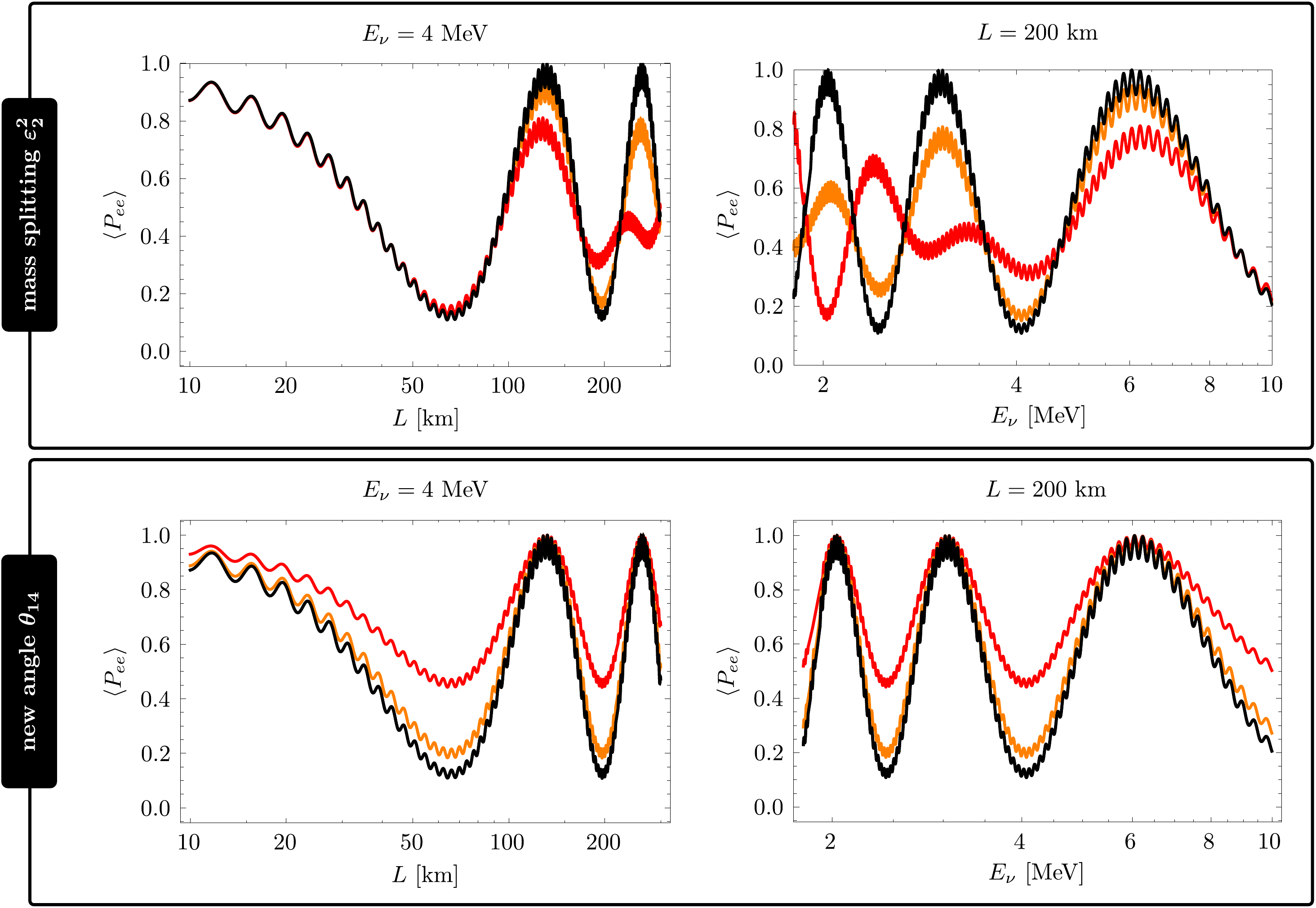}
		\end{centering}
		\protect\caption{\label{fig:PeeRe} Electron neutrino survival
			probability for quasi-Dirac neutrinos with a fixed energy
			$E_{\nu}=4$ MeV as a function of distance (left), and for fixed
			distance $L=200$ km as function of $E_{\nu}$ (right). The standard
			3-generation neutrino oscillation parameters have been fixed at
			their best fit point values \cite{deSalas:2017kay}, to
			which a small perturbation has been added. In the top row, we show
			the effect of mass splittings: $\varepsilon_2^2=0$ (black),
			$\varepsilon_2^2= 10^{-5}$ eV$^2$ (orange) and $\varepsilon_2^2=2\times
			10^{-5}$ eV$^2$ (red). In the bottom row, it is possible to see the
			effect of introducing a non-standard angle: $\theta_{14}=0$ (black),
			$\theta_{14}=\pi/8$ (orange) and $\theta_{14}=\pi/4$ (red).  The
			exact definition of $\varepsilon_2$ and $\theta_{14}$ will be given
			latter in section \eqref{sect:param}.}
	\end{figure}
	\par
\end{center}

A word on nomenclature. The terminologies {\em quasi-Dirac} and {\em
	pseudo-Dirac} neutrinos appear nearly interchangeably in the
literature. We prefer to define quasi-Dirac (QD) neutrinos as being a
mixture of active and sterile states, in contrast with pseudo-Dirac
(PD) neutrinos \footnote{This distinction is based on two early papers
	on the subject \cite{Wolfenstein:1981kw,Valle:1982yw}. Wolfenstein
	\cite{Wolfenstein:1981kw} discussed pairs of active neutrinos, which
	almost preserve lepton number (due to a relative CP-sign) if the
	mixing angle between them is close to maximal and the mass
	splitting is small. He called such particles ``pseudo-Dirac''
	neutrinos. Near the end of the paper, Wolfenstein then extended the
	terminology to mass matrices which contain both, active and sterile
	states. In \cite{Valle:1982yw}, on the other hand, Valle proposed to
	use the terminology ``quasi-Dirac'' neutrinos for active-sterile
	pairs, to differentiate them from ``pseudo-Dirac'' (active-active
	pairs).} which are composed of active states only. In both cases,
the structure of mass and mixing matrices must be such that the lepton
sector is close to preserving one or more $U(1)$ symmetries.

With this definition, quasi-Dirac and pseudo-Dirac neutrinos are then
very different objects, both theoretically and phenomenologically. Let
us briefly mention that various aspects of pseudo-Dirac neutrinos have
been considered in the literature: Magnetic moments and double beta
decay \cite{Petcov:1982ya}, possible mass textures
\cite{Doi:1983wu,Dutta:1994wz,Joshipura:2000ts}, and oscillatory
behavior \cite{Bilenky:1983wt,Bilenky:1987ty,Giunti:1992hk,Nir:2000xn}.
We note in passing that models of pseudo-Dirac neutrinos require
neutrino mass matrices which no longer fit the solar and atmospheric
neutrino oscillation data \cite{Frampton:2001eu,He:2003ih,Brahmachari:2001rn}.
\footnote{PD neutrinos have mass matrices with entries close to zero
	on the diagonal. This is similar to the case discussed in
	\cite{Frampton:2001eu,He:2003ih,Brahmachari:2001rn} for the Zee
	model \cite{Zee:1980ai}.}

Many more papers discussed the phenomenology of quasi-Dirac
neutrinos. For example, double beta decay was first discussed in this
context in \cite{Valle:1982yw}, while \cite{Geiser:1998mr} and
\cite{Krolikowski:1999an,Balaji:2001fi} consider quasi-Dirac neutrinos
as a possible explanation of the atmospheric and solar neutrino
problems, respectively. More ambitiously, explaining atmospheric,
solar and LSND neutrino oscillations simultaneously was discussed in
\cite{Ma:1995gf,Goswami:2001zg}. However, all these proposals are by
now ruled out experimentally, since they predict too much oscillations
into sterile neutrinos. Limits on quasi-Dirac neutrino parameters, on
the other hand, have been derived from solar neutrino data
\cite{deGouvea:2009fp} as well as from solar, atmospheric neutrino data and
cosmology \cite{Cirelli:2004cz}. Furthermore, in
\cite{Beacom:2003eu,Esmaili:2009fk,Esmaili:2012ac,Joshipura:2013yba}
QD neutrinos have been discussed in the context of neutrino
telescopes, such as IceCube.

Quasi-Dirac neutrino oscillations were also discussed in
\cite{Kobayashi:2000md}, where it was claimed that to leading order in
$m_{R,L}/m_{D}$ the flavor composition of the mass eigenstates does
not change (only mass splittings appear), hence oscillations for $n=3$
pairs of quasi-Dirac neutrinos are described by the standard mixing
matrix. This assertion was taken to be true by others
\cite{Beacom:2003eu,Esmaili:2009fk,Esmaili:2012ac,Ahn:2016hhq}, yet we
want to stress that this claim is not correct, as can be seen
from the eqs. (\ref{eq:2})--(\ref{eq:4}). Already for one
generation, these expressions show that the mass splitting and the
departure from maximal mixing are both linearly dependent on $m_{R,L}$ and,
more importantly, they are controlled by orthogonal combinations of
these two parameters. As such, it is even possible to have no mass
splittings at all and at the same time have arbitrary mixing angles.

There are also a number of more theoretical papers discussing how
quasi-Dirac neutrinos could arise.  One possibility is the so-called
``singular'' seesaw where the mass matrix for the singlet neutrinos
($N^c$) has a determinant equal or close to zero
\cite{Stephenson:2004wv}.  Quasi-Dirac neutrinos from such a singular
seesaw with additional type-II seesaw contributions have been
discussed in \cite{McDonald:2004qx}.  Another possibility
\cite{Ahn:2016hhq} involves introducing additional singlets ($S$), as
it is done for the inverse seesaw mechanism \cite{Mohapatra:1986bd}. A
double seesaw is then responsible for producing very light $S$ states
which, together with the active states, form quasi-Dirac neutrinos
\cite{Ahn:2016hhq}.  The authors of \cite{Chang:1999pb} use a Dirac
seesaw to explain the necessary smallness of the Dirac neutrino mass
terms first, and then generate quasi-Dirac states by the addition of a
very small seesaw type-II term. The ``mirror world'' model of
\cite{Berezinsky:2002fa} is another way to obtain these particles.

In models with extended gauge groups quasi-Dirac neutrinos can also
appear. An example is the $E_6$ inspired $331$ model of
\cite{Sanchez:2001ua}.  Here, several electroweak triplets of the
gauge group $SU(3)_L$ are needed to accomodate the Standard Model
leptons, and the observed active light neutrinos are automatically
quasi-Dirac states \cite{Fonseca:2016xsy}. A very different idea,
based on supergravity has been discussed in \cite{Abel:2004tt}. There
it was pointed out that if neutrino Dirac terms are generated from the
K\"ahler potential (instead of the superpotential), neutrinos would be
quasi-Dirac, since Majorana terms come from higher order K\"ahler
potential terms and thus are expected to be suppressed. This idea
\cite{Abel:2004tt} is particularly attractive, since it could, at
least in principle, explain the observed smallness of the Dirac
neutrino mass terms.

In addition to $n$ active neutrinos, models of Dirac neutrinos require
the introduction of $n$ Weyl spinors transforming trivially under the
electroweak gauge group. For this reason, the study of quasi-Dirac
neutrinos necessarily has some overlap with the physics of sterile
neutrinos. Many experiments have searched for sterile neutrinos.  Most
famously, the SNO neutral current measurement rules out dominant
contributions of sterile neutrinos to the solar neutrino oscillations
\cite{Ahmad:2002jz}.  Super-Kamiokande searched for steriles in
atmospheric neutrinos \cite{Abe:2014gda}. OPERA
\cite{Agafonova:2015neo}, MINOS and DayaBay \cite{Adamson:2016jku},
IceCube\cite{Aartsen:2017bap} and NO$\nu$A \cite{Adamson:2017zcg}
published searches for sterile neutrinos. For a more complete list of
references see the recent reviews
\cite{Abazajian:2012ys,Gariazzo:2015rra}. Note, however, that
constraints on sterile are usually derived assuming best fit point
values for the standard oscillation parameters, to which two new
parameters (one angle and one mass splitting) are added in the
fit. This approach does not cover the general quasi-Dirac neutrino
parameter space. In particular, keeping the standard neutrino
parameters fixed can lead to misleading conclusions about limits for
the new/extra parameters. 

There are also some hints for the existence of sterile neutrinos.
However, all these hints point to a new and {\em much larger} mass
scale in oscillations, i.e. $\Delta m^2 \simeq {\cal O}$(1) eV$^2$.
Since these indications imply masses and mixings very different from
those of the standard oscillations, they can not be explained by
quasi-Dirac neutrinos. We thus do not discuss these hints any further
and refer only to the recent review \cite{Gariazzo:2015rra}.

The rest of this paper is organized as follows. In section
(\ref{sect:param}) we discuss the basics of quasi-Dirac oscillations,
constructing general expressions for the mixing matrix for the three
generation case. In section (\ref{sect:exp}) we discuss constraints on
the new, non-standard parameters from various neutrino experiments.
Constraints on quasi-Dirac mass splittings are discussed in section
(\ref{subsect:splits}), while in section (\ref{subsect:angs}) we
discuss the constraints on angles, for the case in which mass
splittings are negligible.  We then close with a short summary and
discussion.

\section{Definitions for quasi-Dirac neutrino oscillations}
\label{sect:param}

Dirac neutrinos can be described either in the weak or in the mass
basis. The two pictures are equivalent. We will choose the latter
one. Consider then a lepton-number preserving model with three active
and three sterile neutrinos ($\nu$ and $N^{c}$).\footnote{Fields with
	no flavor indices should be seen as vectors.} In the basis where the
charged lepton mass matrix is diagonal, the relevant part of the
Lagrangian reads
\begin{align}
	\mathscr{L} & =\frac{g}{\sqrt{2}}\overline{\ell_{L}}\mathds{1}\gamma^{\mu}\nu W_{\mu}^{-}+\nu^{T}m_{\nu}N^{c}+\textrm{h.c.}\;\;\textrm{\ensuremath{\left[\textrm{flavor basis}\right]}}\,.
\end{align}
In order to diagonalize the matrix $m_{\nu}$, both active and sterile
neutrinos must be rotated, $\nu\rightarrow V\nu$ and $N^{c}\rightarrow
V_{N}N^{c}$, such that $m_{\nu}^{\textrm{(diag)}}=V^{T}m_{\nu}V_{N}$:
\begin{align}
	\mathscr{L} & =\frac{g}{\sqrt{2}}\overline{\ell_{L}}V\gamma^{\mu}\nu W_{\mu}^{-}+\nu^{T}m_{\nu}^{\textrm{(diag)}}N^{c}+\textrm{h.c.}\;\;\textrm{\ensuremath{\left[\textrm{mass basis 1}\right]}}\,.
\end{align}
Strictly speaking, the neutrino mass matrix is not yet diagonal since
it is still mixing different states (active and sterile neutrinos).
This can be solved by rewriting $\nu_{i}$ and $N_{i}^{c}$ ($i=1,2,3$)
as $\psi_{i}\equiv1/\sqrt{2}\left(\nu_{i}+N_{i}^{c}\right)$, and
$\psi_{i+3}\equiv i/\sqrt{2}\left(-\nu_{i}+N_{i}^{c}\right)$:
\begin{align}
	\mathscr{L} & =\frac{g}{\sqrt{2}}\overline{\ell_{L}}\Omega\gamma^{\mu}\psi W_{\mu}^{-}+\sum_{j=1}^{6}m^\psi_{j}\psi_{j}\psi_{j}+\textrm{h.c.}\;\;\textrm{\ensuremath{\left[\textrm{mass basis 2}\right]}}\,.
\end{align}
where the masses and the $3\times6$ mixing matrix $\Omega$ have a
special form ($V$ is a $3\times3$ square matrix):
\begin{align}
	m^\psi_{j} & =\left(m_{1},m_{2},m_{3},m_{1},m_{2},m_{3}\right)\;\;\hspace{1.8cm}\textrm{\ensuremath{\left[\textrm{Dirac limit}\right]}}\,,\label{eq:DiracLimit1}\\
	\Omega & =\frac{1}{\sqrt{2}}\left[\begin{array}{ccc}
		\ddots &  & \iddots\\
		& V\\
		\iddots &  & \ddots
	\end{array},\begin{array}{ccc}
	\ddots &  & \iddots\\
	& iV\\
	\iddots &  & \ddots
\end{array}\right]\;\;\textrm{\ensuremath{\left[\textrm{Dirac limit}\right]}}\,.\label{eq:DiracLimit2}
\end{align}

If the pattern of masses and mixing in eqs. (\ref{eq:DiracLimit1})
and (\ref{eq:DiracLimit2}) is perturbed, neutrinos are no longer Dirac
particles and lepton number is violated.  Note that this is equivalent to
switching on the lepton number violating masses $m_{L}$ and $m_{R}$ in
eq. (\ref{eq:MassMat}). We shall now look into the possible departures
from the Dirac limit as seen from the mass basis.

In the case of masses, it is possible to split the three pairs of
$\left(m^\psi_{i},m^\psi_{i+3}\right)$, hence we may introduce three
$\varepsilon_{i}$ such that
\begin{align}
	\left(m_{i}^{\psi}\right)^{2},\left(m_{i+3}^{\psi}\right)^{2}
	& \rightarrow m_{i}^{2}-\frac{\varepsilon_{i}^{2}}{2},m_{i}^{2}+\frac{\varepsilon_{i}^{2}}{2}\,,\label{eq:mass_splittings}
\end{align}
with the understanding that, for quasi-Dirac neutrinos, the
$\varepsilon_{i}$ are small in comparison to the atmospheric and solar
mass scales.  In total there are now five mass parameters relevant for
oscillation experiments: the usual $\Delta m^2_{\rm Atm}$ and $\Delta m^2_{\odot}$,
plus three new $\varepsilon_{i}$ mass splittings. (As usual, the
overall mass scale of neutrinos does not enter the oscillation
probabilities.)

Let us now turn our attention to a generic mixing matrix $\Omega$ with
dimensions $n\times m$. Such a matrix can be described by $2nm$ real
numbers, yet orthonormality of rows
($\Omega\Omega^{\dagger}=\mathds{1}$) imposes $n^{2}$ conditions on
them, and furthermore it is possible to absorb $n$ phases into the
charged lepton fields, hence there is a total of
$n\left(2m-n-1\right)$ real physical degrees of freedom in
$\Omega$. For a $3\times6$ matrix, this corresponds to 12 angles and
12 phases, but note that 5 of these phases cannot be observed in
neutrino oscillation experiments (they correspond to column phases).
The matrix $\Omega$ can be explicitly parametrized as follows
\cite{Schechter:1980gr} (called below the SV parametrization). First,
consider an elementary rotation in the $\left(i,j\right)$ entries
given by the complex number
$\widetilde{\theta}_{ij}\equiv\theta_{ij}\exp i\phi_{ij}$ such that,
in the (1,2) case, it has the form
\begin{align}
	R\left(\widetilde{\theta}_{12}\right) & =\left(\begin{array}{cccc}
		\cos\theta_{12} & -e^{i\phi_{12}}\sin\theta_{12} & 0 & \cdots\\
		e^{-i\phi_{12}}\sin\theta_{12} & \cos\theta_{12} & 0 & \cdots\\
		0 & 0 & 1 & \cdots\\
		\vdots & \vdots & \vdots & \ddots
	\end{array}\right)\,.\label{eq:elementaryRotation}
\end{align}
In the SV parametrization, the $i$-th row of $\Omega$ ($\equiv\Omega_{i}$)
is then given by the expression
\begin{align}
	\Omega_{SV,i}^{T} & =\prod_{a=1}^{i}\prod_{b=a+1}^{6}R\left(\widetilde{\theta}_{ab}\right)e^{(i)}\,,\label{eq:Omega_i}
\end{align}
where $e^{(i)}$ is a column vector with entries $e_{j}^{(i)}=\delta_{ij}$.
We do not give here $\Omega_{SV}$ in full because the expression
is very lengthy.

For a particular arrangement of the 24 angles and phases in eq.
(\ref{eq:Omega_i}), $\Omega$ takes the special form (\ref{eq:DiracLimit2})
which is associated with the Dirac limit. Note that, as usual, one
can write the $3\times3$ square matrix $V$ with three angles and
one phase:
\begin{align}
	V & =\left(\begin{array}{ccc}
		1 & 0 & 0\\
		0 & \cos\theta_{23} & \sin\theta_{23}\\
		0 & -\sin\theta_{23} & \cos\theta_{23}
	\end{array}\right)\left(\begin{array}{ccc}
	\cos\theta_{13} & 0 & \sin\theta_{13}e^{-i\delta}\\
	0 & 1 & 0\\
	-\sin\theta_{13}e^{i\delta} & 0 & \cos\theta_{13}
\end{array}\right)\left(\begin{array}{ccc}
\cos\theta_{12} & \sin\theta_{12} & 0\\
-\sin\theta_{12} & \cos\theta_{12} & 0\\
0 & 0 & 1
\end{array}\right)\,.\label{eq:V_usual}
\end{align}
Unfortunately, it is very complicated to describe the Dirac limit in
the SV parametrization. Hence we make a small modification by
introducing the following $6\times6$ rotation matrix:
\begin{align}
	\Omega\left(\theta_{ij},\phi_{ij}\right) & \equiv\Omega_{SV}\left(\theta_{ij},\phi_{ij}\right)U\;,\,U=\frac{1}{\sqrt{2}}\left(\begin{array}{cc}
		\mathbb{1} & i\mathbb{1}\\
		\mathbb{1} & -i\mathbb{1}
	\end{array}\right)\,.\label{eq:newParametrization}
\end{align}
With this definition, the mixing matrix in eq.
(\ref{eq:DiracLimit2}), with $V$ parametrized as in
eq. (\ref{eq:V_usual}), corresponds to
$\Omega\left(\theta_{ij},\phi_{ij}\right)$, as in
(\ref{eq:newParametrization}), with
$\theta_{i4}=\theta_{i5}=\theta_{i6}=0$ ($i=1,2,3$),
$\phi_{12}=\phi_{23}=0$ and $\phi_{13}=\delta$. In other words, with
this definition the Dirac limit for $\Omega$ simply corresponds to
keeping only the standard three generation neutrino mixing angles
non-zero.

We can then write the probability of neutrino oscillation from a flavor
$\alpha$ to a flavor $\beta$ for an energy $E$ and after a length $L$
as:\footnote{This is true as long as the rows of $\Omega$ are
	orthonormal, i.e.  $\Omega\Omega^{\dagger}=\mathds{1}$.}
\begin{align}
	P\left(\nu_{\alpha}\rightarrow\nu_{\beta}\right) &
	= \left|\sum_{j=1}^{6}\Omega_{\beta j}\Omega_{\alpha j}^{*}
	\exp\left(-\frac{im_{j}^{2}L}{2E}\right)\right|^{2}\label{eq:Posc}
\end{align}
Note that this expression is insensitive to column rephasings
$\Omega\rightarrow\Omega\,\textrm{diag}
\left(e^{i\kappa_{1}},e^{i\kappa_{2}},e^{i\kappa_{3}},e^{i\kappa_{4}},e^{i\kappa_{5}},e^{i\kappa_{6}}\right)$.
It is easy to show that eq. (\ref{eq:Posc}) reduces to the standard
oscillation formula in the Dirac limit.

\section{Current experimental limits and future prospects}
\label{sect:exp}

As discussed in the previous section, the full parameter space for a
system of 3 pairs of QD neutrinos has 30 free parameters: Two
independent $\Delta m_{ij}^2$ plus one overall mass scale, three
$\varepsilon_i^2$, twelve angles and twelve phases. Even discounting the
five Majorana phases and the overall mass scale, which can not be
probed in oscillation experiments, the remaining number of parameters
is much too large to fit simultaneously.

Nearly all experimental data, on the other hand, is consistent with
the standard picture of only three active neutrino species
participating in oscillations \cite{deSalas:2017kay}, i.e.  two mass
squared differences ($\Delta m^2_{\rm Atm}$ and $\Delta m^2_{\odot}$),
three mixing angles ($\theta_{23}$, $\theta_{12}$ and
$\theta_{13}$) plus one phase ($\delta$) are sufficient to describe the
data. As mentioned in the introduction, there are also some hints for
sterile neutrinos with a mass scale of the order of $\Delta m^2 \sim
{\cal O}$(eV) \cite{Abazajian:2012ys,Gariazzo:2015rra}.  However, all
these hints are at most of the order of ($2-3$) $\sigma$, we will thus
not take them into account in the following. Instead, since the
standard three generation picture seems to describe the data well, we
will consider ``small'' perturbations and derive limits on particular
combinations of non-standard parameters.

In order to deal effectively with the large number of parameters
controlling quasi-Dirac neutrino oscillations, we will consider two
simplified scenarios:
\begin{enumerate}
	\item First, we take one non-zero $\varepsilon_{i}$ at a time.  In
	these one-parameter extensions, very stringent limits on
	$\varepsilon_{i}$ are found, in agreement with earlier analysis, see
	for example \cite{deGouvea:2009fp,Cirelli:2004cz}. We then extend
	this analysis to two new parameters: One mass splitting plus one new
	angle. This second step allows us to identify ``blind spots'' in the
	oscillation experiments, i.e. degenerate minima in particular
	directions in parameter space, where limits on mass splittings are
	much worse than in the one parameter fits. We then discuss a
	particular parametrization of these degenerate directions in
	parameter space, where the effects of $\varepsilon_{i}$ can be
	decoupled from oscillation experiments nearly completely.
	\item In the second setup, we discuss the limit where mass splittings
	are too small to be measured in oscillation experiments, hence there
	are just angles and phases of $\Omega$ to deal with. In this
	situation, it can be shown that from the 24 parameters in $\Omega$
	only 13 combinations enter the oscillation probabilities of active
	neutrinos. Moreover, since there is only very limited information on
	oscillations involving $\nu_{\tau}$, we can in practice restrict
	ourselves to experiments involving $\nu_{e}$'s and $\nu_{\mu}$'s.
	There are then only 7 combinations of the 24 angles $\theta_{ij}$
	and phases $\phi_{ij}$ which appear in the oscillation
	probabilities. We discuss the construction of these 7 quantities,
	the current constraints and possible tests for quasi-Dirac neutrinos
	in this limit.
\end{enumerate}

In our analysis we do not take into account the data from every
existing oscillation experiment. Given the scarcity of data on $\tau$
neutrinos, we ignored it altogether, concentrating instead on the
available charged current data for $e$ and $\mu$ neutrinos and
anti-neutrinos.  Also, we focus on those experiments, which should
provide the most important constraints for quasi-Dirac
neutrinos. First, we consider KamLAND \cite{Abe:2008aa} since it fixes
most accurately the so-called solar mass splitting $\Delta
m^2_{\odot}$. From the solar neutrino experiments we fit Super-K
elastic scattering data \cite{Abe:2016nxk} and from Borexino the
measured $pp$ \cite{Bellini:2014uqa} and $^{7}Be$ fluxes
\cite{Bellini:2011rx}.  This choice is motivated by the fact that
Super-K \cite{Abe:2016nxk} has the most accurate data in the high
energy range, while \cite{Bellini:2014uqa,Bellini:2011rx} fix the low
energy part of the solar neutrino spectrum.

We take the atmospheric neutrino data from \cite{Kajita:2014koa}. We
concentrate in our fit on the sub-GeV sample, since the lowest
energetic neutrinos will be most sensitive to small values of
$\varepsilon_i^2$, as demonstrated in fig. (\ref{fig:PeeAtmE3}). In
addition, we use data from the MINOS collaboration (muon and anti-muon
neutrino survival) \cite{Adamson:2013whj} and T2K (muon neutrino
survival and muon to electron neutrino transition) \cite{Abe:2015awa},
since these two experiments determine $\Delta m^2_{\rm Atm}$ better
than atmospheric data. And, finally, we take into account data from
DayaBay \cite{An:2016ses}, since from the three current reactor
neutrino experiments DayaBay determines $\theta_{13}$ with the
smallest error.

\begin{center}
	\begin{figure}[tbph]
		\begin{centering}
			\includegraphics[scale=0.6]{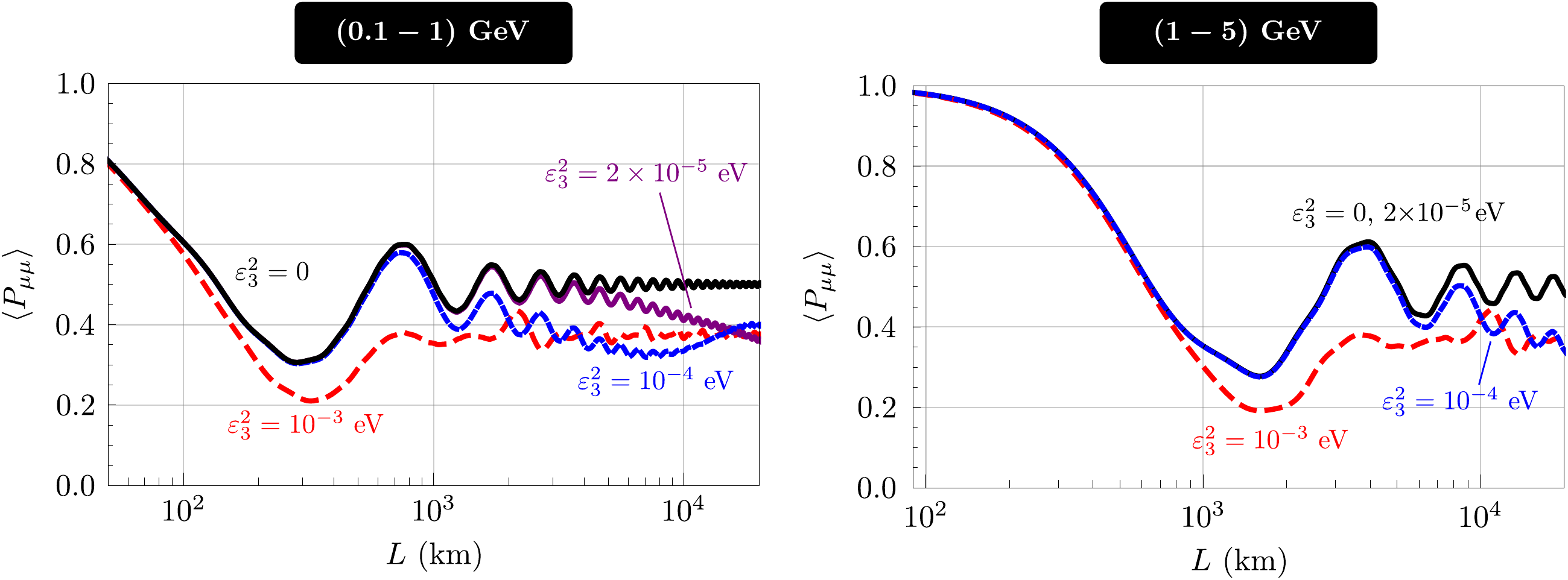}
		\end{centering}
		\protect\caption{\label{fig:PeeAtmE3} Averaged atmospheric muon
			neutrino survival probability for neutrinos with energies
			$E_{\nu}=(0.1-1)$ GeV (left) and $E_{\nu}=(1-5)$ GeV (right), as a
			function of distance ($L$), for different choices of
			$\varepsilon_3^2$. Lower neutrino energies are more sensitive to
			small $\varepsilon_3$ values. This plot is calculated with the
			simplifying assumptions of $\sin^2\theta_{23}=1/2$, $\theta_{13}=0$
			and $\Delta m^2_{\odot}=0$.}
	\end{figure}
	\par
\end{center}

In our fits, we use a simple $\chi^2$ method to determine the allowed
ranges of model parameters. We take statistical and systematic errors
from the experimental publications, to which we added a further
(small) systematic error for the uncertainties in our theoretical
calculations. This latter systematic error was chosen such that our
simulations reproduce the allowed parameter ranges for the standard
oscillation parameters, determined by the respective experiment,
within typically (1-1.5) $\sigma$ c.l. ranges. Note that we do not
attempt to do a precision global fit for standard neutrino oscillation
parameters.  Rather, we consider reproducing the experimental results
for the standard case as a test for the reliability of our derived
limits.

\subsection{Limits on mass splittings $\varepsilon_{i}$ }
\label{subsect:splits}

\subsubsection{One parameter limits}

We will first discuss limits derived on $\varepsilon_{i}$ assuming one
$\varepsilon_{i} \ne 0$ at a time and taking all non-standard angles
to be zero. Table (\ref{Tab:EpsLim}) shows limits on
$\varepsilon_i^2$ and the corresponding experimental data sets used to
derive the limits.

\begin{table}
	\begin{center}
		\begin{tabular}{cccc}
			\toprule
			Experiment& $\varepsilon_1^2$ [eV$^2$] &$\varepsilon_2^2$  [eV$^2$] &
			$\varepsilon_3^2$ [eV$^2$] \tabularnewline
			\midrule
			KamLAND & $7.7 (3.4) \times 10^{-6}$ & $1.7 (1.0) \times 10^{-5}$ & -- \tabularnewline
			Solar + KamLAND & $1.7 (1.3)\times 10^{-11}$ & $1.7 (1.5) \times 10^{-11}$ & -- \tabularnewline
			DayaBay + MINOS + T2K
			& -- & $1.5 (0.9) \times 10^{-4}$ & $1.3 (0.074) \times 10^{-3}$ \tabularnewline
			Super-K + DayaBay + MINOS + T2K
			& -- & $1.9 (1.8) \times 10^{-5}$& $1.2 (1.1) \times 10^{-5}$ \tabularnewline
			JUNO & $1.7 (0.07) \times 10^{-5}$ & $2.3 (0.09) \times 10^{-5}$
			& $6.0 (2.2) \times 10^{-5}$\tabularnewline
			\bottomrule
		\end{tabular}
	\end{center}
	\caption{ \label{Tab:EpsLim} 95 \% upper limits on $\varepsilon_i^2$
		derived from different experimental data sets. Two numbers are given
		for each case; the first one is the limit obtained marginalizing
		over two standard oscillation parameters (see text), the second (in
		brackets) is the limit obtained for the best fit point value of the
		standard oscillation parameters. For a discussion see text.}
\end{table}

For each case listed in table (\ref{Tab:EpsLim}), we have calculated
the upper limits on the $\varepsilon_i^2$ twice: (a) marginalizing
over two of the standard neutrino oscillation parameters and (b) for
the best fit point value of the standard parameters. Marginalization
over standard oscillation parameters leads to less stringent limits.
However, the importance of this marginalization procedure differs
widely for different experiments.  For example, in the case of
KamLAND, bounds on $\varepsilon_1^2$ of the order of roughly $10^{-5}$
are derived marginalizing over the allowed ranges of $\Delta
m^2_{\odot}$ and $\sin^2\theta_{12}$, while for the best fit values
of these last two parameters, the limits are more stringent by
``only'' roughly a factor 2. 

\begin{figure}[tbph]
	\begin{center}
		\includegraphics[scale=0.65]{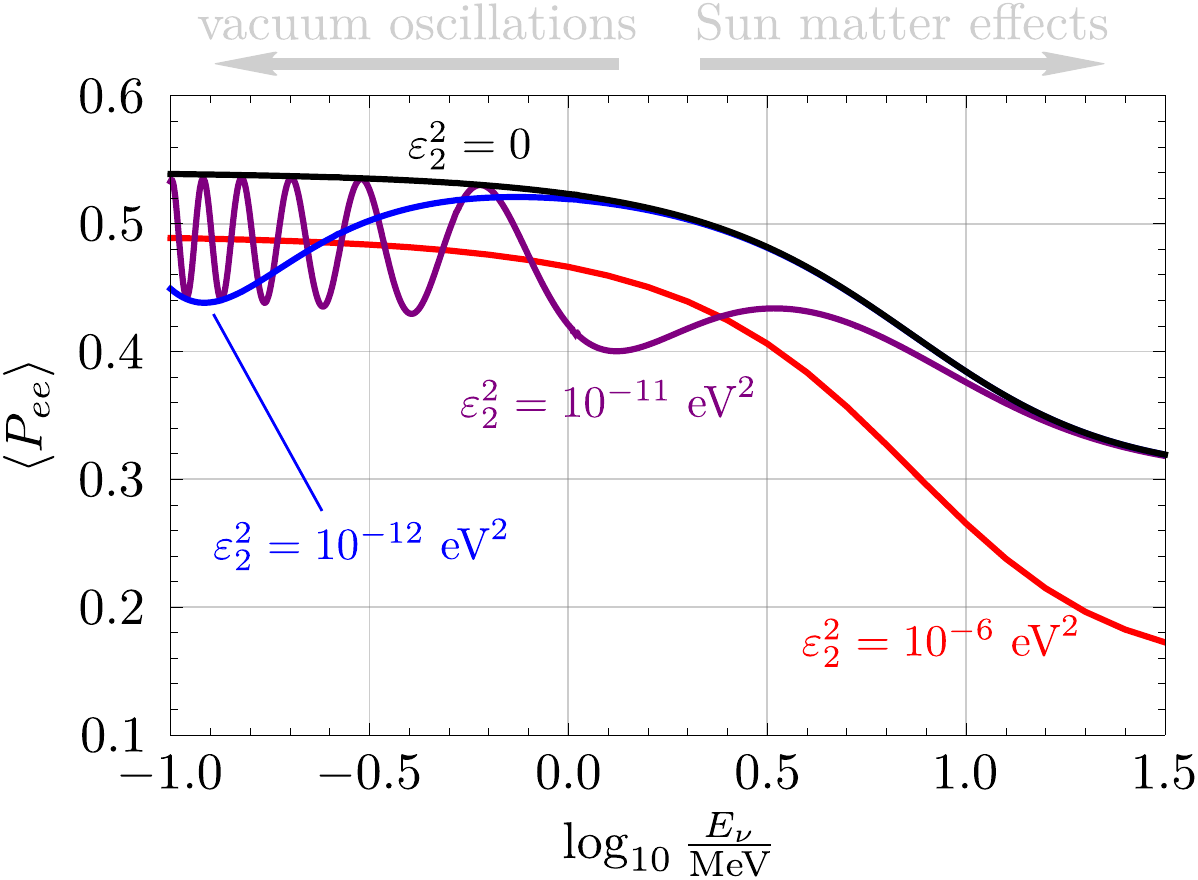}
	\end{center}
	\protect\caption{\label{fig:PeeSolE2} Solar neutrino survival
		probability as a function of neutrino energy, for different choices
		of $\varepsilon_2^2$. Solar angle and mass splitting have
		been fixed at their best fit values in this plot
		\cite{deSalas:2017kay}.}
\end{figure}

As the table shows, the strongest constraints on $\varepsilon_1^2$ and
$\varepsilon_2^2$ come from solar neutrino data. This is easily
understood from fig. (\ref{fig:PeeSolE2}), which shows the electron
neutrino survival probability as a function of neutrino energy for
different values of $\varepsilon_2^2$ and for the best fit values of
the standard solar oscillation parameters. For low values of neutrino
energies, vacuum oscillations dominate and so very small
$\varepsilon_2^2$ can be probed up to a scale essentially determined
by the Earth-Sun distance ($\sim10^{-12,-11}\textrm{ eV}^{2}$). Note
that a non-zero $\varepsilon_2^2$ reduces $P_{ee}$, but this reduction
could be hidden in the relatively large error bar of the low-energy
measurements.\footnote{Due to the annual variation of the Earth-Sun
	distance, for values of $\varepsilon_2^2$ larger than
	$\sim10^{-10,-9}\textrm{ eV}^{2}$, the oscillations are averaged
	over, so only an overall reduction of survival probability is seen.}
Nevertheless, at higher neutrino energies, a similar reduction of
$P_{ee}$ is produced due to matter effects in the sun.  Since Super-K
data provides a very accurate measurement of $P_{ee}$ at these higher
energies, one can rule out values of $\varepsilon_2^2$ which can not
be excluded by the Borexino measurements alone. The situation is very
similar for $\varepsilon_1^2$: limits on $\varepsilon_1^2$ and
$\varepsilon_2^2$ are then of the order of $10^{-11}$ eV$^2$ if we
take the best fit values of the standard solar oscillation parameters;
slightly less stringent numbers are obtained when marginalizing over
the standard parameter uncertainties.

Solar neutrino experiments have essentially no sensitivity to
$\varepsilon_3^2$. This is simply due to the smallness of
$\theta_{13}$ ($\sin^2\theta_{13} \simeq 0.0215$
\cite{deSalas:2017kay}).  Thus, we have to rely on experiments testing
the atmospheric scale to derive limits on $\varepsilon_3^2$.
Table (\ref{Tab:EpsLim}) quotes numbers for two cases.

In the first scenario, we have combined data from DayaBay
\cite{An:2016ses}, T2K \cite{Abe:2015awa} and MINOS
\cite{Adamson:2013whj}: DayaBay fixes most accurately $\theta_{13}$,
while both MINOS and T2K measure $\Delta m^2_{\rm Atm}$ with rather
small errors. Here, the limit on $\varepsilon_2^2$ is (not
surprisingly) less stringent than the one derived from KamLAND (or
solar).  Depending on whether or not $\Delta m^2_{\rm Atm}$ and
$\sin^2\theta_{\rm 23}$ are used at their best fit value or
marginalized over, we get very different limits on $\varepsilon_3^2$.
This is due to the fact that when scanning over the standard
oscillation parameters, the $\chi^2$ function has two almost
degenerate minima: one for small values of $\varepsilon_3^2$ and
another for $\varepsilon_3^2$ of the order $10^{-3}$ eV$^2$.  However,
as the table also shows (second case), this non-standard solution is
excluded, once we add Super-K atmospheric neutrino sub-GeV data to the
fit. With the combination of these four experiments limits on
$\varepsilon_2^2$ and $\varepsilon_3^2$ are again of order $10^{-5}$
eV$^2$.

In the last line of table (\ref{Tab:EpsLim}) we give our forecast of
the sensitivity of the planned experiment JUNO \cite{An:2015jdp}.
JUNO will measure $\Delta m^2_{\odot}$ and $\Delta m^2_{\rm Atm}$ very
precisely and thus it will also be able to derive limits on any
$\varepsilon_i^2$.  However, our results indicate that, despite being
a very precise experiment, JUNO will not lead to a major improvement
over existing limits on $\varepsilon_i^2$. Here, it is important to
stress that limits using the best fit point and limits marginalizing
over standard parameter uncertainties are very different. This can be
traced back again to a near-degeneracy in the $\chi^2$-function: For
$\epsilon_i^2$ of the order of $\Delta m^2_{ij}$ one has two only
slightly different oscillation lengths contributing in the fit, which
can give a better description than a single oscillation length.

In summary, strong limits on mass splittings can be derived from
atmospheric and solar neutrino data ($\varepsilon_3^2 \sim
{\cal O}(10^{-5})$ eV$^2$ and $\varepsilon_{1,2}^2 \sim
{\cal O}(10^{-11})$ eV$^2$, respectively) in the case where no other
extra parameter is added to the standard neutrino oscillation picture.

\subsubsection{Two parameter case}

While the discussion in the previous subsection seems to show that
constraints on QD mass splittings are very stringent, we will now see
that this conclusion is valid only under the assumption that no other
non-standard parameter is different from zero.

As a simple example, consider the electron neutrino survival
probability at distances short enough that the effects of $\Delta
m^2_{\odot}$ can be neglected.\footnote{To a good approximation, this
	is the situation in the DayaBay experiment.}  We shall consider the
particular example of a non-zero $\varepsilon_3^2$ and a non-zero
$\theta_{16}$ angle, defined in section \ref{sect:param}. One
finds that
\begin{equation}\label{eq:pee2p}
	P_{ee} = 1 - 2 c_{13}^2c_{16}^2
	\Big( \Delta m_{ee}^- (s_{16}+c_{16}s_{13})^2
	+\Delta m_{ee}^+ (s_{16}- c_{16}s_{13})^2 \Big)\,,
\end{equation}
where $c_{ij}$ and $s_{ij}$ are short-hands for $\cos\theta_{ij}$
and $\sin\theta_{ij}$ and
\begin{equation}\label{eq:Delee}
	\Delta m_{ee}^{\pm} \equiv c_{12}^2\sin^2\left[\frac{L}{4E}\left(\Delta m_{31}^{2}\pm\varepsilon_{3}^{2}/2\right)\right]  
	+  s_{12}^2\sin^2\left[\frac{L}{4E}\left(\Delta m_{32}^{2}\pm\varepsilon_{3}^{2}/2\right)\right]\,.
\end{equation}
It is straightforward to see that the above expression for the
neutrino survival probability remains (nearly) unchanged if we swap
$\theta_{13}$ by $\theta_{16}$. This is true up to very small terms
proportional to $\Delta(P_{ee}) \propto (\Delta m_{ee}^--\Delta
m_{ee}^+) (c_{13}-c_{16}) s_{13}s_{16}$. More specifically, in the
limit where $\Delta m_{ee}^{\pm}$ have the same value, $P_{ee}$ is
only a function of the combination
$\sin^2\theta_{13}+\sin^2\theta_{16}$.  Thus, there will be a
near-degeneracy of the relevant $\chi^2$ function involving these two
angles, and so values (or limits) derived for one of these parameters, 
without varying the other, will be misleading.

There is, however, another more interesting degeneracy associated to
eq. (\ref{eq:pee2p}).  In calculating this expression we have used a
certain parametrization for the mass splitting, which we may call the
{\it symmetric parametrization}: $m_i,m_{i+3} \to
\sqrt{m_i^2-\varepsilon_i^2/2}, \sqrt{m_i^2+\varepsilon_i^2/2}$.
Choosing $\sin\theta_{13}=\tan\theta_{16}$, the second term inside the
bracket in eq. (\ref{eq:pee2p}) vanishes (this choice corresponds to
$\Omega_{e6}=0$).  So, by adjusting $\Delta m^2_{31}$ and
$\varepsilon_3^2$ we can keep $\Delta m^2_{31}-\varepsilon_3^2/2$
constant and equal to the best fit point value of $\Delta m^2_{\rm
	Atm}$, in which case there will be no upper limit on
$\varepsilon_3^2$ itself coming from the electron neutrino survival
experiments.

\begin{figure}[tbph]
	\begin{center}
		\includegraphics[scale=0.4]{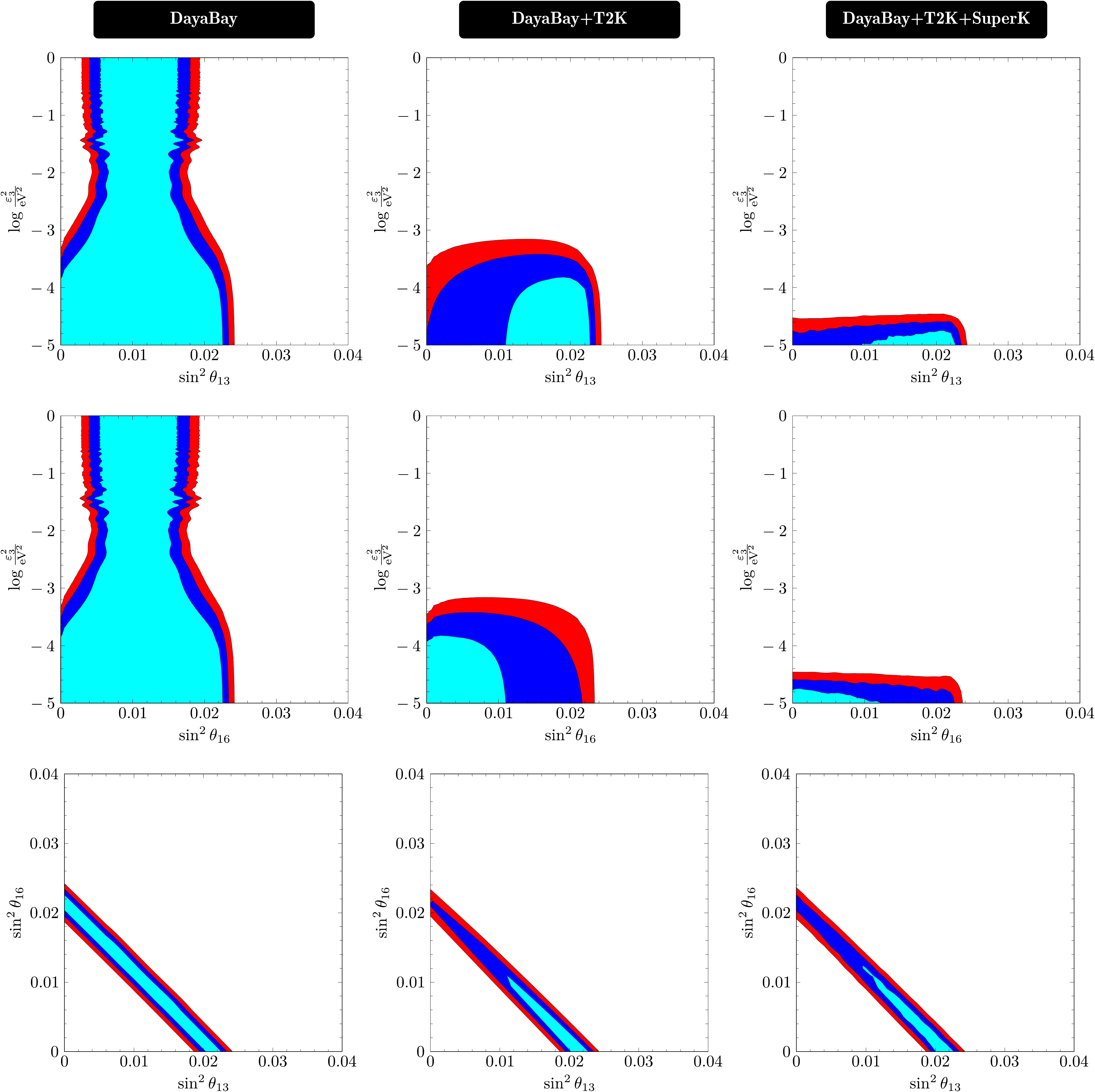}
	\end{center}
	\protect\caption{\label{fig:DegTh16} Allowed parameter ranges for
		$\varepsilon_3^2$, $\sin^2\theta_{13}$ and $\sin^2\theta_{16}$ for
		different experiments. The parameter planes always marginalize over
		the parameter not shown and all calculations used the best fit point
		value for $\Delta m^2_{\rm Atm}$.  In the plots on the left only
		DayaBay data is taken into account; the middle panel combines
		DayaBay with T2K and the panel to the right shows the combination of
		DayaBay, T2K with Super-K atmospheric neutrino data. The different
		coloured regions present the 1, 2 and 3 $\sigma$ c.l. allowed
		regions (cyan, blue and red). For discussion see text.}
\end{figure}

Note that we could have defined $m_i,m_{i+3} \to
m_i,\sqrt{m_i^2+\varepsilon_i^2}$.\footnote{Numerically this leads to
	the same limits on $\varepsilon_i^2$, as long as the mass splitting
	is much smaller than the relevant $\Delta m_{ij}^2$ (i.e., the solar
	or atmospheric scale).}  We call this the {\it asymmetric
	parametrization}. Rewriting eq. (\ref{eq:pee2p}) with this
parametrization, the first term inside the bracket would not depend on
$\varepsilon_3^2$ at all, so it becomes obvious that for the choice of
$\sin\theta_{13}=\tan\theta_{16}$ all dependence of $P_{ee}$ on
$\varepsilon_3^2$ disappears.  Fig. (\ref{fig:DegTh16}) shows these
parameter degeneracies in the space
($\varepsilon_3^2,\sin^2\theta_{13},\sin^2\theta_{16}$), using only
the DayaBay data (on the left column). The underlying scan was done in
the asymmetric parametrization, which is numerically simpler to
implement. The plot in the upper and middle panel show clearly that
there is no upper limit on $\varepsilon_3^2$ in this scan. The lower
plot shows the degeneracy in parameter space under the exchange of
$\theta_{13} \leftrightarrow \theta_{16}$.

We can break this particular degeneracy in parameter space, by adding
more experiments. T2K measures two probabilities: (a) The muon
neutrino survival probability, $P_{\mu\mu}$, and (b) the electron
neutrino appearance probability, $P_{\mu e}$, both at values of $L/E$
which give access to the atmospheric neutrino mass scale, $\Delta
m^2_{\rm Atm}$. If the only non-standard angle different from zero is
$\theta_{16}$, then $P_{\mu\mu}$ will not depend on $\theta_{16}$ at
all, while $P_{\mu e}$ will have $\theta_{16}$-dependence which is
different from the one of $P_{ee}$. Thus, adding T2K data to the scan
is enough to break the degeneracy in $\theta_{13} \leftrightarrow
\theta_{16}$ hence an upper limit on $\varepsilon_3^2$ reappears. The
middle column of fig. (\ref{fig:DegTh16}) illustrates this point; it
shows the results of a combined scan over $\varepsilon_3^2$,
$\sin^2\theta_{13}$ and $\sin^2\theta_{16}$ for DayaBay plus T2K data.
By comparison of the right with the middle column of
fig. (\ref{fig:DegTh16}), one can clearly see that the addition of
Super-K data generates a strong upper limit on $\varepsilon_3^2$, for
this particular choice of parameter subspace.

Given these results, one might wonder if there are particular
directions in parameter space for which oscillation experiments become
completely blind to QD neutrino mass splittings. Recall that the
blind (or: degenerate) direction discussed above for $P_{ee}$ corresponds to
the particular choice of $\Omega_{e6}=0$. In a similar way, for
example, $P_{\mu\mu}$ would loose any sensitivity to $\varepsilon_3^2$
if $\Omega_{\mu6}=0$. Thus, with some special choice of $\theta_{16}$
and $\theta_{26}$ such that both $\Omega_{e6}$ and $\Omega_{\mu6}$ are
zero at the same time, one can indeed make DayaBay and T2K blind to
variations of $\varepsilon_3^2$.

While it is possible, in principle, to calculate the combination of
angles $\theta_{ij}$ and phases $\phi_{ij}$ (defined in section
\eqref{sect:param}) associated to these blind directions, in the
following we will consider a simpler alternative.  Consider a unitary
rotation of the columns $i$ and $i+3$ of the mixing matrix
$\Omega$. Since we are not interested in column phases, such a
rotation is governed by just two parameters ($\varphi_{i}$ and
$\beta_i$):
\begin{align}
	\left(\begin{array}{cc}
		\vdots & \vdots\\
		\rotatebox{90}{\;\;\;col. $i$ of $\Omega$} & \rotatebox{90}{col. $i+3$ of $\Omega$\;}\\
		\vdots & \vdots
	\end{array}\right) & \rightarrow \left(\begin{array}{cc}
	\vdots & \vdots\\
	\rotatebox{90}{\;\;\;col. $i$ of $\Omega$} & \rotatebox{90}{col. $i+3$ of $\Omega$\;}\\
	\vdots & \vdots
\end{array}\right)\left(\begin{array}{cc}
\cos\varphi_{i} & e^{i\beta_{i}}\sin\varphi_{i}\\
-e^{-i\beta_{i}}\sin\varphi_{i} & \cos\varphi_{i}
\end{array}\right)\label{eq:NoMassSplittingsDegeneracy}
\end{align}
Now, recall that in the Dirac limit (see eq.  (\ref{eq:DiracLimit2}))
the columns $i$ and $i+3$ of the mixing matrix $\Omega$ are
proportional to each other,
$\Omega_{\alpha,i}=-i\Omega_{\alpha,i+3}=V_{\alpha,i}/\sqrt{2}$.  This
means that applying the $\left(\varphi_{i},\beta_{i}\right)$
transformation to the Dirac neutrino mixing matrix and, without loss of
generality setting $\beta_i=0$, we obtain:
\begin{align}
	\left(\begin{array}{cc}
		\vdots & \vdots\\
		\Omega_{\alpha,i} & \Omega_{\alpha,i+3}\\
		\vdots & \vdots
	\end{array}\right) & =\left(\begin{array}{cc}
	\vdots & \vdots\\
	\frac{\cos\varphi_{i}+\sin\varphi_{i}}{\sqrt{2}}V_{\alpha,i} & i\frac{\cos\varphi_{i}-\sin\varphi_{i}}{\sqrt{2}}V_{\alpha,i}\\
	\vdots & \vdots
\end{array}\right)\, .
\end{align}
From the last of these equations it can be seen that the $i$'th column
(the ($i+3$)'th column) of $\Omega$ vanishes, if one chooses
$\varphi_{i} = 3 \pi/4$ ($\varphi_{i} =\pi/4$).

\begin{center}
	\begin{figure}[tbph]
		\begin{center}
			\includegraphics[scale=0.5]{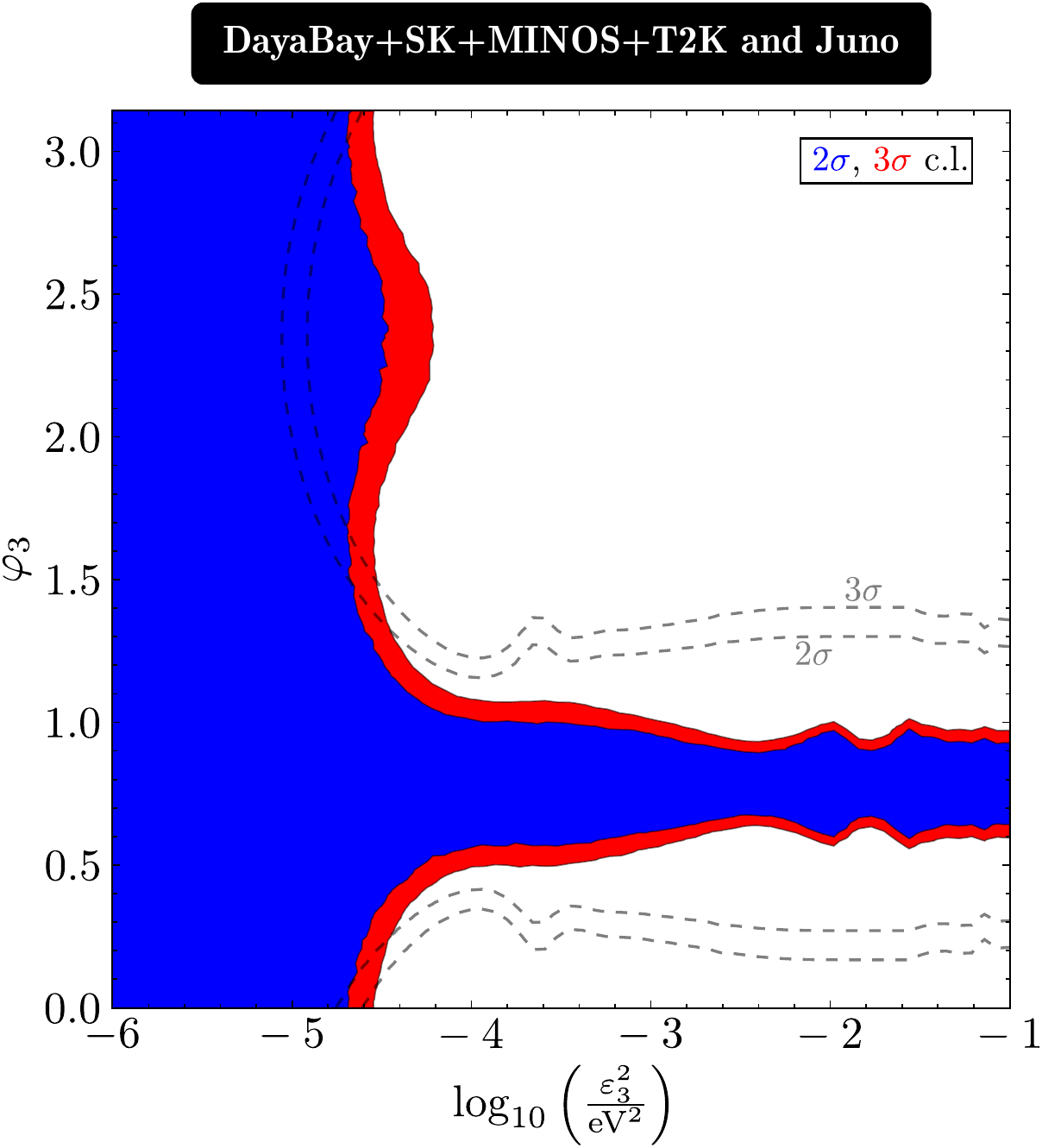}
		\end{center}
		\protect\caption{\label{fig:AtmPhi36} Allowed parameter space in the
			plane ($\varepsilon_3^2,\varphi_{3}$) using DayaBay, T2K, MINOS and
			Super-K atmosheric neutrino data. The coloured plane shows the 2 and
			3 $\sigma$ c.l. allowed regions (blue and red). The dashed lines
			show the expected limits for JUNO. The
			asymmetric parametrization of the mass splitting
			($m_3,\sqrt{m_3^2+\varepsilon_3^2}$) was used, so for $\varphi_{3}=\pi/4$
			there is no sensitivity to $\varepsilon_3^2$.}
	\end{figure}
	\par
\end{center}

Fig. (\ref{fig:AtmPhi36}) shows a scan over the allowed parameter
space in the plane ($\varepsilon_3^2,\varphi_{3}$) using DayaBay, T2K,
MINOS and Super-K atmospheric neutrino data. In agreement with the
above discussion, there is a blind spot where no limit on
$\varepsilon_3^2$ exist. This blind direction corresponds to the
choice of $\varphi_{3}=\pi/4$.\footnote{Shifting $\varepsilon_3^2$ one
	could alternatively define ($\sqrt{m_3^2+\varepsilon_3^2},m_3$). In
	that case, the blind spot occurs at $\varphi_{3}=3 \pi/4$ instead.}
Fig. (\ref{fig:AtmPhi36}) also shows that the addition of JUNO data
can lead only to a marginally improved limit.

We now turn to a discussion of $\varepsilon_1^2$ and
$\varepsilon_2^2$.  For these two parameters, again solar neutrino
physics provides the most important constraints. As above for
$\varepsilon_3^2$, we can define a rotation angle $\varphi_{1}$
($\varphi_{2}$) between the columns 1 and 4 (2 and 5) of the mixing
angle which will mitigate the effects of a non-zero $\varepsilon_1^2$
($\varepsilon_2^2$). Fig. (\ref{fig:PeeSolPhi25}) shows the $P_{ee}$
probability for solar neutrinos as a function of neutrino energy for
two different values of $\varepsilon_2^2$ and various values of
$\varphi_{2}$.\footnote{This probability is averaged over the
	variations of the Earth-Sun distance, and neutrino production point
	inside the Sun.} The results for $\varepsilon_1^2$ and $\varphi_{1}$
are completely analogous. As the figure shows, for $\varphi_{2}=\pi/4$
again the effects of $\varepsilon_2^2$ completely decouple from the
oscillation probability.

\begin{center}
	\begin{figure}[tbph]
		\begin{centering}
			\hskip5mm\includegraphics[scale=0.6]{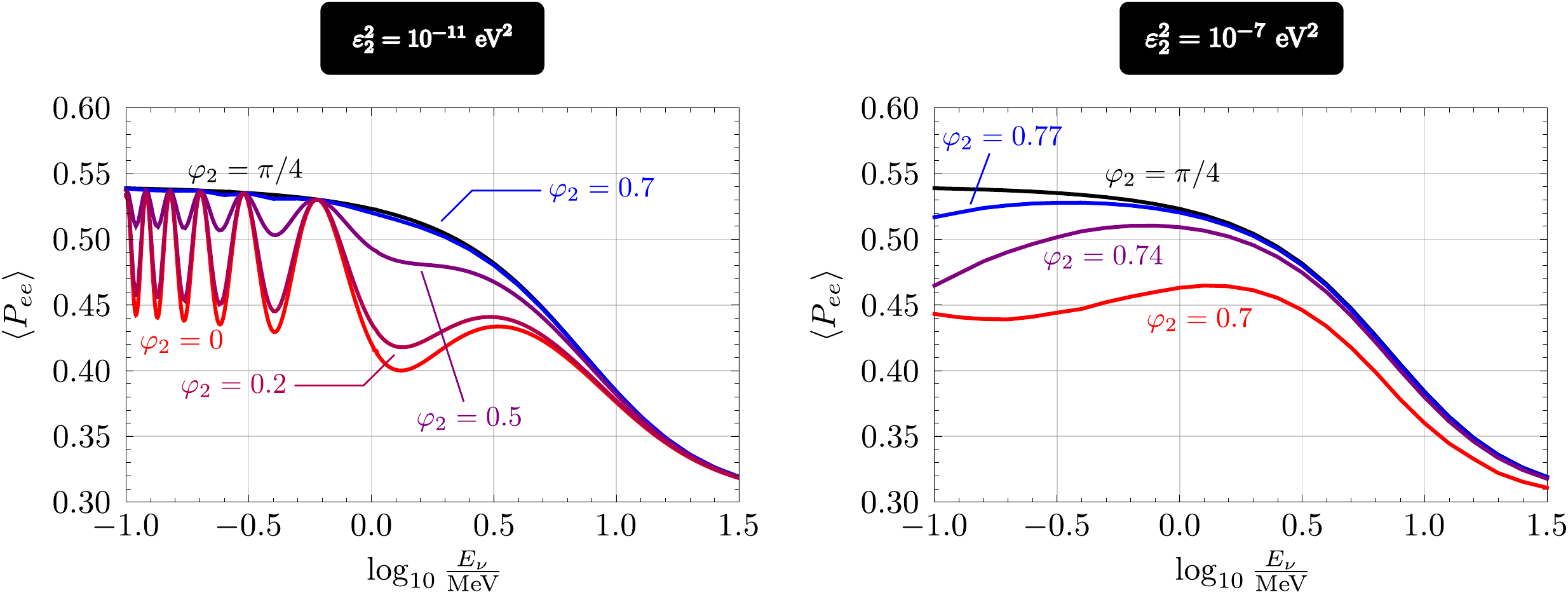}
		\end{centering}
		\protect\caption{\label{fig:PeeSolPhi25} Average solar neutrino
			survival probability as a function of neutrino energy, for different
			choices of $\varphi_{2}$ and two different values of
			$\varepsilon_2^2$ and $\varphi_2$.}
	\end{figure}
	\par
\end{center}

\begin{figure}[tbph]
	\begin{center}
		\includegraphics[scale=0.5]{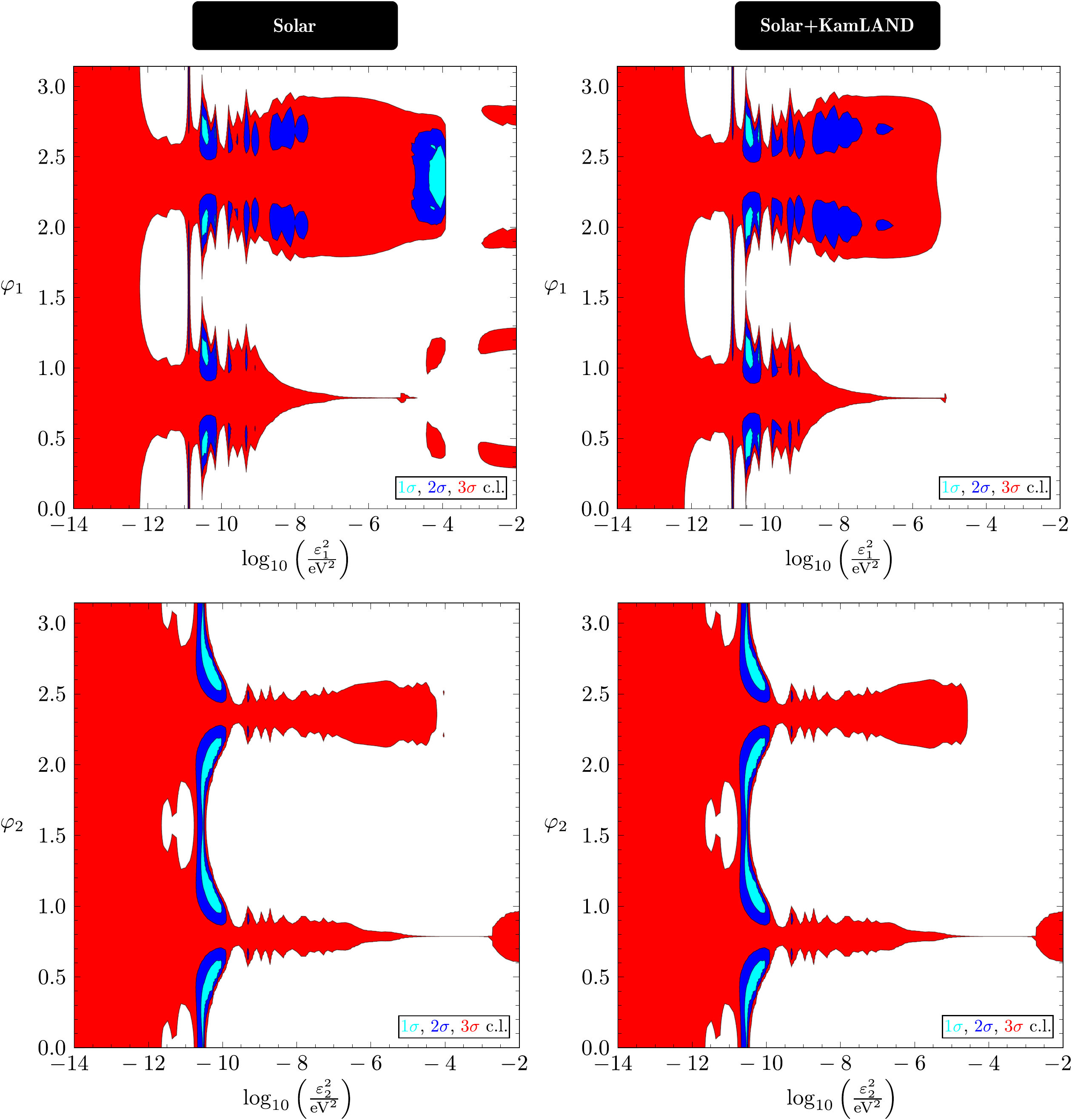}
	\end{center}
	\protect\caption{\label{fig:ChiSolPhi14} Allowed parameter range in
		the space $\varphi_{1},\varepsilon_1^2$ (top) and
		$\varphi_{2},\varepsilon_2^2$ (bottom). To the left: Solar data, to
		the right solar data + KamLAND.  This plot uses the best fit point
		values for $\Delta m^2_{\odot}$ and $\sin^2\theta_{12}$ from the
		global fit. This combination of data shows a slight preference for a
		non-zero value of the mass splitting, for a discussion see text.}
\end{figure}

Fig. (\ref{fig:ChiSolPhi14}) shows the allowed parameter space in the
two planes ($\varphi_{1},\varepsilon_1^2$) and ($\varphi_{2},\varepsilon_2^2$)
using solar data and combining solar data with KamLAND.  These plots
have been calculated using the best fit point values for $\Delta m^2_{\odot}$
and $\sin^2\theta_{12}$ from the global fit \cite{deSalas:2017kay}.
The plots show in all cases that there exists a slight preference,
between (2-2.5) $\sigma$ in all cases, for non-zero values of
$\varepsilon_i^2$. Note that the preferred solution of the solar data
in the region of $\varphi_{1} \sim 3/4 \pi$ and $\varepsilon_1^2 \sim
(10^{-4.5}-10^{-4})$ is ruled out by KamLAND. However, even combining
solar and KamLAND data some preference for non-zero $\varepsilon_i^2$
of the order of very roughly $10^{-10.5}$ eV$^2$ remains.

\begin{figure}[tbph]
	\begin{center}
		\includegraphics[scale=0.5]{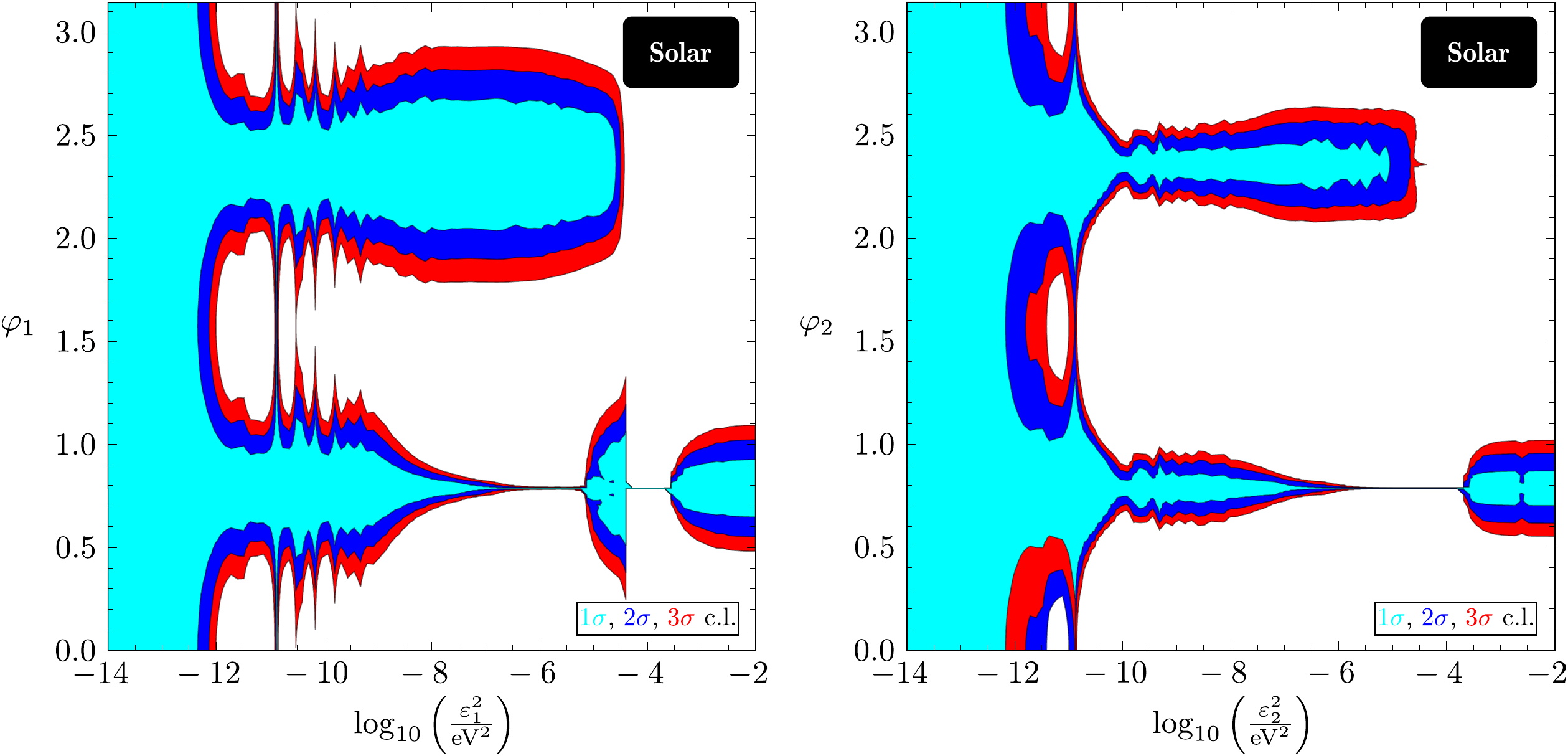}
	\end{center}
	\protect\caption{\label{fig:PeeSolLowDelM} Allowed parameter range in
		the space $\varphi_{1},\varepsilon_1^2$ (left) and
		$\varphi_{2},\varepsilon_2^2$ (right) from solar data, using $\Delta
		m^2_{\odot} = 4 \times 10^{-5}$ eV$^2$.}
\end{figure}

We have traced back this preference for non-zero mass splittings in
solar data to the well-known difference in the best fit points from
$\Delta m^2_{\odot}$ in solar and KamLAND data. As can be seen also in
the latest global fits \cite{deSalas:2017kay}, solar data prefers a
$\Delta m^2_{\odot}$ around $(4-5) \times 10^{-5}$ eV$^2$, while
KamLAND prefers $\Delta m^2_{\odot} \simeq 7.6 \times 10^{-5}$
eV$^2$. This tension between the two data sets is roughly of the order
of 2 $\sigma$, with the error bar dominated by the larger error on
$\Delta m^2_{\odot}$ in the solar data set. We have therefore
recalculated the constraints on ($\varphi_{1},\varepsilon_1^2$) from
solar data for a value of $\Delta m^2_{\odot} = 4 \times 10^{-5}$
eV$^2$.  Fig. (\ref{fig:PeeSolLowDelM}) shows the results of such a
scan. As can be seen, in this calculation there is no longer any
preference for a non-zero value of $\varepsilon_1^2$.

Note that such a low value of $\Delta m^2_{\odot}$ is ruled out by
many $\sigma$ from the KamLAND data. Thus, a small non-zero mass
splitting could provide, in principle, a solution for the observed
tension between solar and KamLAND data. 

In summary, by introducing one mass splitting at a time, we extracted
bounds for the pairs of parameters $\left(\varepsilon_{i},\varphi_{i}\right)$,
$i=1,2,3$. In the limit where $\varphi_{i}$ is $\left(2\pm1\right)/4\pi$,
one column of the mixing matrix vanishes and therefore the mass splitting
$\varepsilon_{i}$ becomes unobservable. For this reason, one expects
that for reasonably large values of $\varepsilon_{i}$ there must
be tight limits on $\left|\varphi_{i}-\left(2\pm1\right)/4\pi\right|$,
meaning that $\varphi_{i}$ has to be quite far from the Dirac limit
($\varphi_{i}=0$). On the other hand, for a small enough value of
the mass splitting $\varepsilon_{i}$, the associated oscillation
length eventually become larger than the baseline of the relevant
experiments, and in that case $\varphi_{i}$ becomes unconstrained.

\subsection{Quasi-Dirac neutrinos in the limit $\varepsilon_{i}\to 0$}
\label{subsect:angs}

If masses are degenerate as in eq. (\ref{eq:DiracLimit1}), then the
oscillation probability formula will not change under unitary
rotations of the columns $i$ and $i+3$ of the mixing matrix, see
eq. \eqref{eq:NoMassSplittingsDegeneracy}. In other words,
\begin{align}
	\left(\begin{array}{cc}
		\vdots & \vdots\\
		\rotatebox{90}{\;\;\;col. $i$ of $\Omega$} & \rotatebox{90}{col. $i+3$ of $\Omega$\;}\\
		\vdots & \vdots
	\end{array}\right) & \rightarrow \left(\begin{array}{cc}
	\vdots & \vdots\\
	\rotatebox{90}{\;\;\;col. $i$ of $\Omega$} & \rotatebox{90}{col. $i+3$ of $\Omega$\;}\\
	\vdots & \vdots
\end{array}\right)U(i)\label{eq:NoMassSplittingsDegeneracy2}
\end{align}
for unitary matrices $U(i)$ ($i=1,2,3$) leaves
$P\left(\nu_{\alpha}\rightarrow\nu_{\beta}\right)$ unchanged.  Hence,
there is a $U(2)^{3}$ redundancy in our description of $\Omega$, and
in turn this means that out of the 24 parameters describing the mixing
matrix, oscillation experiments are only sensitive to 13.\footnote{The
	counting goes as follows: each $U(2)$ describes 4 redundancies in
	the parameters, hence there is a total of 12 redundancies in
	$U(2)^{3}$.  However, one of them corresponds to the irrelevance of
	multiplying $\Omega$ by an overall phase; that was already taken
	care of when row phases were removed from the mixing matrix. Hence
	we are left with 24-12+1 real parameters which affect the neutrino
	oscillation probabilities if no $\varepsilon_{i}$'s are introduced.}
This number is further reduced to 7 if we ignore tau neutrinos. In
this latter case the oscillation probabilities can be written as: 
\begin{align}
	P\left(\nu_{e}\rightarrow\nu_{e}\right) & =1+\left(1-X_{1}-X_{2}\right)X_{2}\mathcal{A}_{12}+\left(1-X_{1}-X_{2}\right)X_{1}\mathcal{A}_{13}+X_{1}X_{2}\mathcal{A}_{23}\,,\\
	P\left(\nu_{\mu}\rightarrow\nu_{\mu}\right) & =1+\left(1-X_{3}-X_{4}\right)X_{4}\mathcal{A}_{12}+\left(1-X_{3}-X_{4}\right)X_{3}\mathcal{A}_{13}+X_{3}X_{4}\mathcal{A}_{23}\,,\\
	P\left(\nu_{e}\rightarrow\nu_{\mu}\right) & =-\left(X_{6}+\textrm{Re}X_{7}\right)\mathcal{A}_{12}-\left(X_{5}+\textrm{Re}X_{7}\right)\mathcal{A}_{13}+\textrm{Re}X_{7}\mathcal{A}_{23}+\textrm{Im}X_{7}\left(\mathcal{B}_{12}-\mathcal{B}_{13}+\mathcal{B}_{23}\right)\,,\label{eq:Pemu_quasiDirac}
\end{align}
with the oscillating factors $\mathcal{A}_{ij}\equiv-4\sin^{2}\left[\left(m_{i}^{2}-m_{j}^{2}\right)L/\left(4E\right)\right]$
and $\mathcal{B}_{ij}\equiv2\sin\left[\left(m_{i}^{2}-m_{j}^{2}\right)L/\left(2E\right)\right]$
and the 7 parameters $X_{i}$ defined as follows: 
\begin{gather}
	X_{1}\equiv\left|\Omega_{e3}\right|^{2}+\left|\Omega_{e6}\right|^{2}\,,\,X_{2}\equiv\left|\Omega_{e2}\right|^{2}+\left|\Omega_{e5}\right|^{2}\,,\label{eq:Xi_1}\\
	X_{3}\equiv\left|\Omega_{\mu3}\right|^{2}+\left|\Omega_{\mu6}\right|^{2}\,,\,X_{4}\equiv\left|\Omega_{\mu2}\right|^{2}+\left|\Omega_{\mu5}\right|^{2}\,,\\
	X_{5}\equiv\left|\Omega_{e3}\Omega_{\mu3}^{*}+\Omega_{e6}\Omega_{\mu6}^{*}\right|^{2}\,,\,X_{6}\equiv\left|\Omega_{e2}\Omega_{\mu2}^{*}+\Omega_{e5}\Omega_{\mu5}^{*}\right|^{2}\,,\\
	X_{7}\equiv\left(\Omega_{e3}\Omega_{\mu3}^{*}+\Omega_{e6}\Omega_{\mu6}^{*}\right)\left(\Omega_{e2}^{*}\Omega_{\mu2}+\Omega_{e5}^{*}\Omega_{\mu5}\right)\,.\label{eq:Xi_2}
\end{gather}

As a side remark, we would like to point out here that a similar approach could, in principle, be used in the presence of
one $\varepsilon_{i}$: in this case, instead of 7, there would be 9 combinations of angles and phases to take into account.\footnote{Consider a non-zero $\varepsilon_{1}$
	(for $\varepsilon_{i=2,3}\neq0$, the changes to the following
	expressions are trivial). Then the
	$P\left(\nu_{e}\rightarrow\nu_{e}\right)$,
	$P\left(\nu_{e}\rightarrow\nu_{\mu}\right)$ and
	$P\left(\nu_{\mu}\rightarrow\nu_{\mu}\right)$ probabilities depend
	only on the following quantities:
	\begin{gather}
		\widehat{X}_{1}\equiv\left|\Omega_{e3}\right|^{2}+\left|\Omega_{e6}\right|^{2}\,,\,\widehat{X}_{2}\equiv\left|\Omega_{e2}\right|^{2}+\left|\Omega_{e5}\right|^{2}\,,\,\widehat{X}_{3}\equiv\left|\Omega_{e1}\right|^{2}\,,\\
		\widehat{X}_{4}\equiv\left|\Omega_{\mu3}\right|^{2}+\left|\Omega_{\mu6}\right|^{2}\,,\,\widehat{X}_{5}\equiv\left|\Omega_{\mu2}\right|^{2}+\left|\Omega_{\mu5}\right|^{2}\,,\,\widehat{X}_{6}\equiv\left|\Omega_{\mu1}\right|^{2}\,,\\
		\widehat{X}_{7}\equiv\left|\Omega_{e3}\Omega_{\mu3}^{*}+\Omega_{e6}\Omega_{\mu6}^{*}\right|^{2}\,,\,\arg\left(\widehat{X}_{8}\right)\equiv\arg\left(\Omega_{e1}\Omega_{\mu1}^{*}\Omega_{e4}^{*}\Omega_{\mu4}\right)\,,\\
		\arg\left(\widehat{X}_{9}\right)\equiv\arg\left[\Omega_{e1}\Omega_{\mu1}^{*}\left(\Omega_{e2}^{*}\Omega_{\mu2}+\Omega_{e5}^{*}\Omega_{\mu5}\right)\right]\,.
	\end{gather}
}

Note that the $X_{i}$ defined above can take any value in our framework, provided
that the following constraints are obeyed:
\begin{enumerate}
	\item the first six $X_{i}$ are non-negative real numbers;
	\item neither $X_{1}+X_{2}$ nor $X_{3}+X_{4}$ can be larger than 1;
	\item $X_{5}\leq X_{1}X_{3}$ and $X_{6}\leq X_{2}X_{4}$;
	\item the norm of $X_{7}$ is fixed by $X_{5}$ and $X_{6}$ ($\left|X_{7}\right|^{2}=X_{5}X_{6}$),
	so even though $X_{7}$ is a complex parameter, only $\arg\left(X_{7}\right)$
	is an independent degree of freedom;
	\item $X_{5}+X_{6}+2\cos\left[\arg\left(X_{7}\right)\right]\sqrt{X_{5}X_{6}}$
	cannot be bigger than $\left(1-X_{1}-X_{2}\right)\left(1-X_{3}-X_{4}\right)$.
\end{enumerate}
These conditions are a consequence of the definitions of the $X_{i}$
and the fact that the rows of the mixing matrix $\Omega$ are orthonormal
($\Omega\Omega^{\dagger}=\mathds{1}$). Taking them into account,
we are able to pick out all valid points in the $X_{i}$ parameter
space, without ever referencing back to specific entries of the mixing
matrix.

For reference, the values of these $X_{i}$ parameters in the Dirac
limit as a function of the standard $\theta_{12}$, $\theta_{13}$,
$\theta_{23}$ and $\delta$ parameters (see eq. (\ref{eq:V_usual}))
are the following:
\begin{gather}
	X_{1}=\sin^{2}\theta_{13}\,,\,X_{2}=\sin^{2}\theta_{12}\cos^{2}\theta_{13}\,,\,X_{3}=\cos^{2}\theta_{13}\sin^{2}\theta_{23}\,,\label{eq:Xi_DiracLimit_1}\\
	X_{4}=\sin^{2}\theta_{12}\sin^{2}\theta_{13}\sin^{2}\theta_{23}+\cos^{2}\theta_{12}\cos^{2}\theta_{23}-\frac{1}{2}\sin2\theta_{12}\sin2\theta_{23}\sin\theta_{13}\cos\delta\,,\\
	X_{5}=X_{1}X_{3}\,,\,X_{6}=X_{2}X_{4}\,,\\
	X_{7}=\sin\theta_{12}\sin\theta_{13}\cos^{2}\theta_{13}\sin\theta_{23}\left(-\sin\theta_{12}\sin\theta_{13}\sin\theta_{23}+e^{-i\delta}\cos\theta_{12}\cos\theta_{23}\right)\,.\label{eq:Xi_DiracLimit_2}
\end{gather}

\begin{figure}[tbph]
	\begin{centering}
		\includegraphics[scale=0.37]{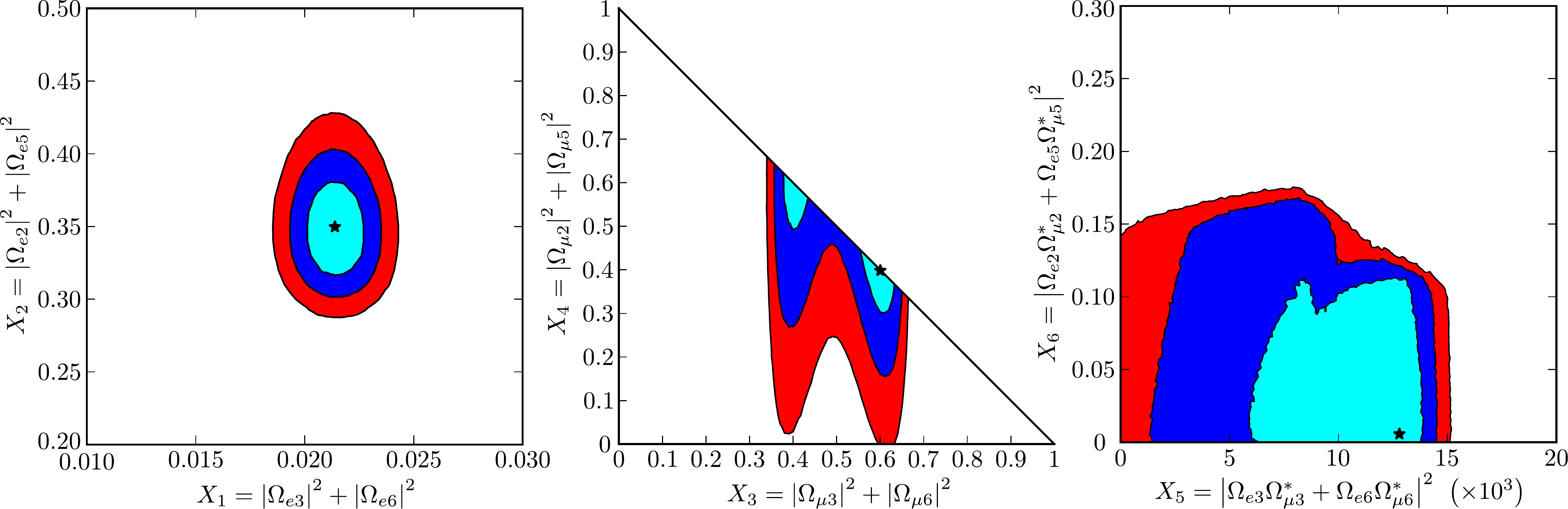}
		\par\end{centering}
	
	\protect\caption{\label{fig:X1X2X3X4X5X6}One, two and three $\sigma$ regions in the
		$\left(X_{1},X_{2}\right)$, $\left(X_{3},X_{4}\right)$ and $\left(X_{5},X_{6}\right)$
		which are allowed by electron neutrino survival data from at KamLAND,
		DayaBay, SuperK and Borexino, muon neutrino survival data at MINOS
		and T2K and muon to electron transition data from T2K. This plots
		were obtained from a 7-dimensional scan of the $X_{i}$ defined in
		eqs. (\ref{eq:Xi_1})--(\ref{eq:Xi_2}) and marginalizing over
		5 variables.}
\end{figure}

\begin{figure}[tbph]
	\begin{centering}
		\includegraphics[scale=0.5]{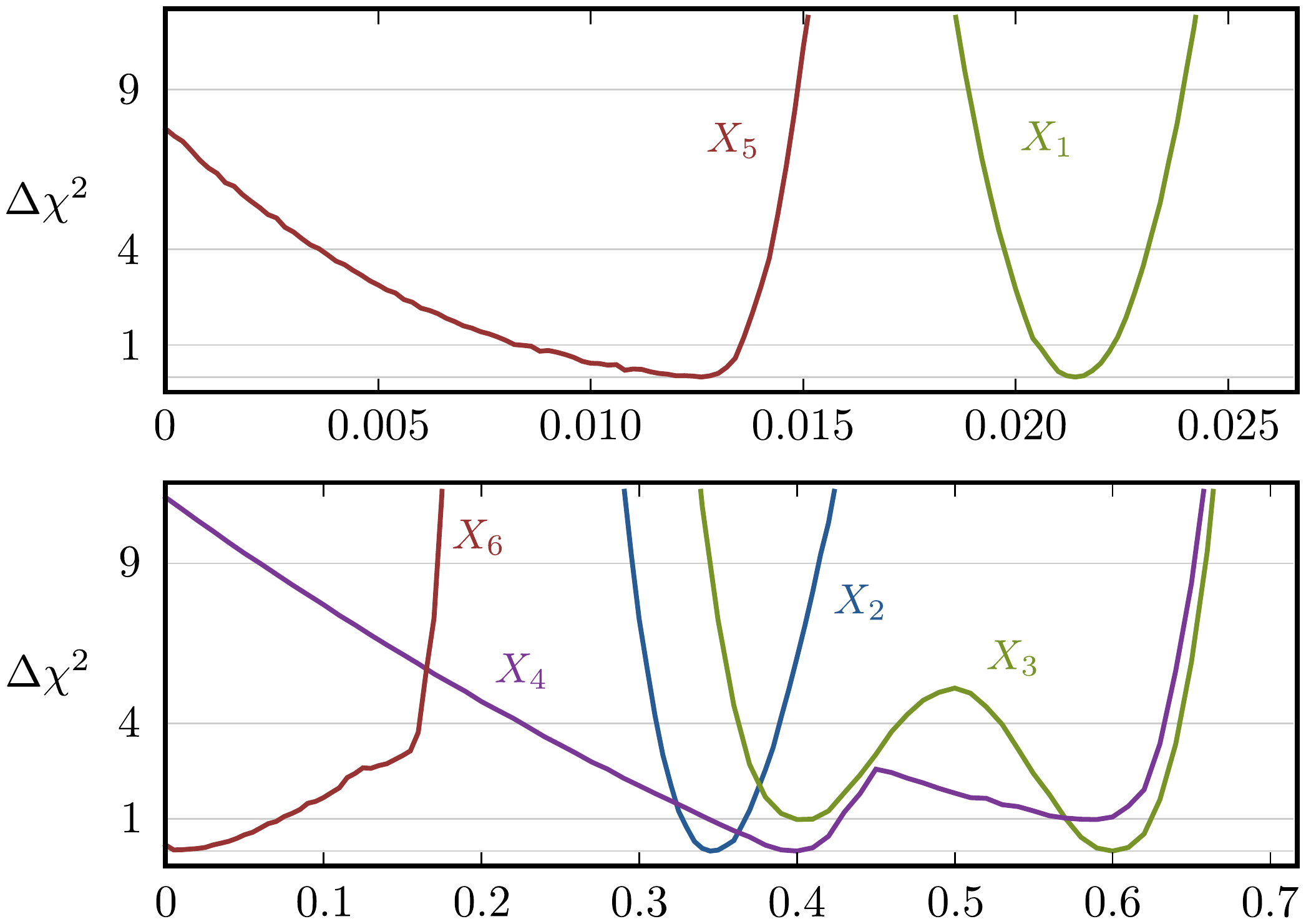}
		\par\end{centering}
	
	\protect\caption{\label{fig:XiMarginalized}Marginalized $\Delta\chi^{2}$ values for
		the variables $X_{1,\cdots,6}$ defined in eqs. (\ref{eq:Xi_1})--(\ref{eq:Xi_2}).
		The $\chi^{2}$ function for $X_{7}$ is essentially flat. These parameters
		are a function of the entries of the mixing matrix only; no mass splittings
		$\varepsilon_{i}$ were considered.}
\end{figure}

Using $\Delta m_{\rm Atm}^{2}=2.55\times10^{-3}$ eV and $\Delta m_{\odot}^{2}=7.56\times10^{-5}$,
we performed a 7-dimensional scan over all $X_{i}$. Electron neutrino
survival data from at KamLAND, DayaBay, SuperK and Borexino was used,
together with muon neutrino survival data at MINOS and T2K and $\nu_{\mu}\rightarrow\nu_{e}$
T2K data. The allowed values in the planes $\left(X_{1},X_{2}\right)$,
$\left(X_{3},X_{4}\right)$ and $\left(X_{5},X_{6}\right)$ are shown
in fig. (\ref{fig:X1X2X3X4X5X6}). The marginalized $\chi^{2}$
function for $X_{1,\cdots,6}$ is shown in fig. (\ref{fig:XiMarginalized});
the marginalized $\chi^{2}\left[\arg\left(X_{7}\right)\right]$ function
is not shown as it is essentially flat from 0 to $2\pi$.

\begin{figure}[tbph]
	\begin{centering}
		\includegraphics[scale=0.5]{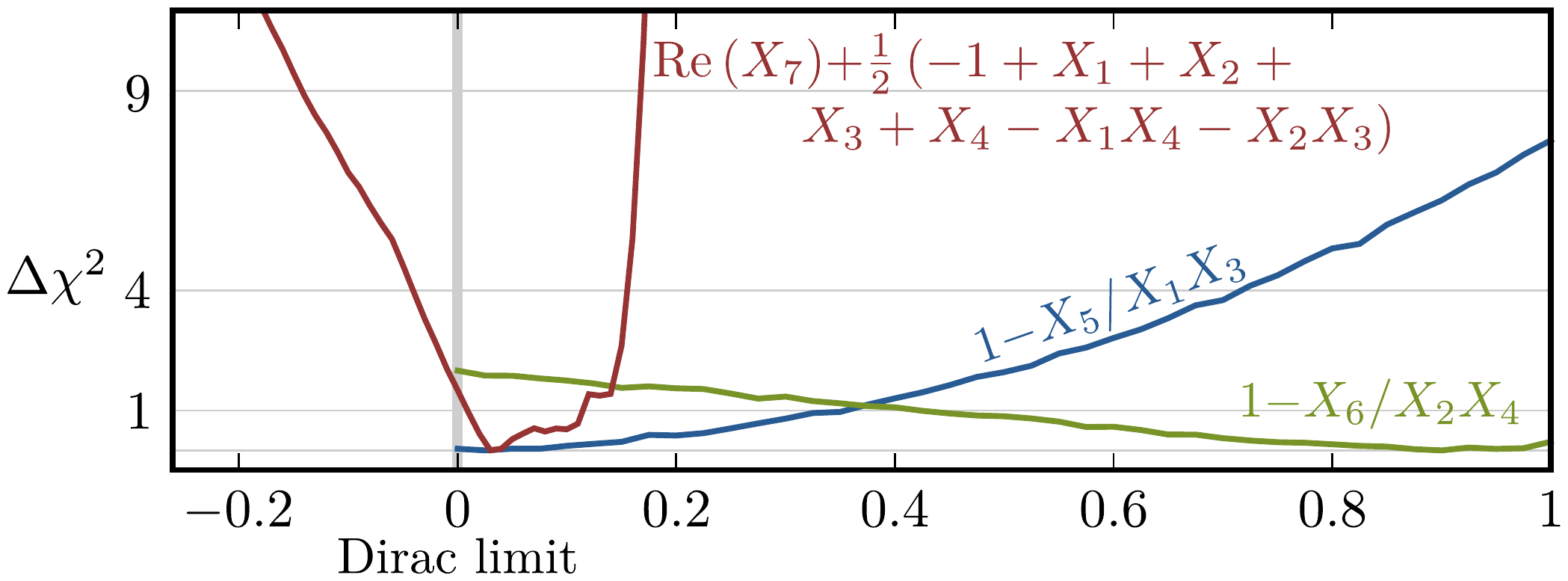}
		\par\end{centering}
	
	\protect\caption{\label{fig:DiracnessXlimits}$\Delta\chi^{2}$
		functions for the three combinations of parameters which, when equal
		to 0 simultaneously, signal the Dirac limit (see
		eq. (\ref{eq:XsDiracnessConditions})).}
\end{figure}

Overall, the bounds on these 7 parameters are broadly consistent with
the standard three neutrino oscillation picture. In other words, by
substituting in expressions
(\ref{eq:Xi_DiracLimit_1})--(\ref{eq:Xi_DiracLimit_2}) the numbers
obtained for $\theta_{12}$, $\theta_{13}$, $\theta_{23}$ and $\delta$
from global fits \cite{deSalas:2017kay}, we get values for the $X_{i}$
roughly in agreement with figs. \eqref{fig:X1X2X3X4X5X6} and
\eqref{fig:XiMarginalized}. In order to see clearly that current data
is consistent with the Dirac limit, note that in this latter case
there are only 4 independent parameters. Thus, it follows that the
standard three neutrino oscillation picture must correspond to three
relations among the 7 $X_{i}$. These are
\begin{gather}
	X_{5}=X_{1}X_{3}\,,\,X_{6}=X_{2}X_{4}\textrm{ and }\textrm{Re}\left(X_{7}\right)=\frac{1}{2}\left(1-X_{1}-X_{2}-X_{3}-X_{4}+X_{1}X_{4}+X_{2}X_{3}\right)\,,\label{eq:XsDiracnessConditions}
\end{gather}
and from fig. (\ref{fig:DiracnessXlimits}) one can see that
oscillation data is compatible with each of these equalities within
$\sim1\sigma$.  The three together are disfavored only at
$\min\left(\chi_{\textrm{Dirac}}^{2}\right)-\min\left(\chi^{2}\right)=1.9$
so, assuming no mass splittings $\varepsilon_{i}$, there is currently
no significant indication for quasi-Dirac neutrinos.

We would like to point out that even in the absence of new mass
scales, quasi-Dirac neutrinos can, in principle, be distinguished from
some other scenarios through oscillation experiments.  In particular,
consider 3 active neutrinos and a non-unitary $3\times3$ mixing matrix
$V$. The oscillation probabilities are then given by the expressions
\begin{align}
	P\left(\nu_{e}\rightarrow\nu_{e}\right) & =F\left(J_{ee}^{12},J_{ee}^{13},J_{ee}^{23}\right)+J_{ee}^{12}\mathcal{A}_{12}+J_{ee}^{13}\mathcal{A}_{13}+J_{ee}^{23}\mathcal{A}_{23}\,,\\
	P\left(\nu_{\mu}\rightarrow\nu_{\mu}\right) & =F\left(J_{\mu\mu}^{12},J_{\mu\mu}^{13},J_{\mu\mu}^{23}\right)+J_{\mu\mu}^{12}\mathcal{A}_{12}+J_{\mu\mu}^{13}\mathcal{A}_{13}+J_{\mu\mu}^{23}\mathcal{A}_{23}\,,\\
	P\left(\nu_{e}\rightarrow\nu_{\mu}\right) & =F^{\prime}\left(J_{ee}^{12},J_{ee}^{13},J_{ee}^{23},J_{\mu\mu}^{12},J_{\mu\mu}^{13},J_{\mu\mu}^{23},J_{e\mu}^{12},J_{e\mu}^{13},J_{e\mu}^{23}\right)+\textrm{Re}\left(J_{e\mu}^{12}\right)\mathcal{A}_{12}+\textrm{Re}\left(J_{e\mu}^{13}\right)\mathcal{A}_{13}\nonumber \\
	& +\textrm{Re}\left(J_{e\mu}^{23}\right)\mathcal{A}_{23}+\textrm{Im}\left(J_{e\mu}^{12}\right)\mathcal{B}_{12}+\textrm{Im}\left(J_{e\mu}^{13}\right)\mathcal{B}_{13}+\textrm{Im}\left(J_{e\mu}^{23}\right)\mathcal{B}_{23}\,,
\end{align}
where $J_{\alpha\beta}^{ij}=V_{\alpha i}^{*}V_{\beta j}^{*}V_{\beta
	i}V_{\alpha j}$.  The exact form of the functions $F$ and $F'$ which
control the 0-distance neutrino behavior is not important for the
present discussion. The more important point is that with a
non-unitary $V$ one can have an oscillatory behavior which is
impossible to reproduce with quasi-Dirac neutrinos, and vice-versa.

For example, the $\mathcal{B}_{ij}$ coefficients in
$P\left(\nu_{e}\rightarrow\nu_{\mu}\right)$ do not need to be related
for a non-unitary $V$, while for quasi-Dirac neutrinos they must be
the same (up to a minus sign --- see
eq. \eqref{eq:Pemu_quasiDirac}). On the other hand, note that from the
$J_{ee}^{ij}$ and $J_{\mu\mu}^{ij}$ one can extract the modulus of the
absolute value of all $J_{e\mu}^{ij}$, hence by measuring
$P\left(\nu_{e}\rightarrow\nu_{e}\right)$ and
$P\left(\nu_{\mu}\rightarrow\nu_{\mu}\right)$, as well as the
coefficients $\mathcal{B}_{ij}$ in
$P\left(\nu_{e}\rightarrow\nu_{\mu}\right)$, the coefficients of the
oscillatory factors $\mathcal{A}_{ij}$ in
$P\left(\nu_{e}\rightarrow\nu_{\mu}\right)$ are fixed for a
non-unitary $V$ (up to $\pm$ signs). However, for quasi-Dirac
neutrinos no such constraint exists. So, with this short theoretical
argument, one can conclude that in principle these two non-standard
neutrino scenarios can be distinguished through oscillation
experiments.

\section{Summary}
\label{sect:Summary}

In general, neutrinos can have lepton number violating (Majorana) and
lepton number conserving (Dirac) mass terms. If the lepton number
violating mass terms are smaller than the lepton number preserving
ones, neutrinos are {\em quasi-Dirac} particles. Phenomenologically,
this corresponds to the existence of three pairs of neutrinos with
slightly different masses, hence oscillation experiments are sensitive
not only to the usual solar and atmospheric mass scales, but also to
three small mass splittings $\varepsilon_i$. Furthermore, for
quasi-Dirac neutrinos there are more than 3+1 angles and phases to be
considered.  In this work, we have analyzed the constraints on these
quasi-Dirac neutrino parameters imposed by current neutrino
oscillation data and also briefly discussed the potential of the
future JUNO experiment to improve upon existing constraints.

In section \eqref{sect:param} we have discussed a fully general
parametrization of the lepton sector for three generations of
quasi-Dirac neutrinos. In addition to the charged lepton masses, there
is a total of 6 masses, 12 angles and 12 phases. Oscillation
experiments are not sensitive to the overall neutrino mass scale nor
to 5 of the phases (which are of the Majorana type). Hence we are left
with a 24-dimensional model space, compared to the six-dimensional
space for an ordinary three generation case ($\Delta m^2_{\odot}$,
$\Delta m^2_{\rm Atm}$, $\theta_{12}$, $\theta_{13}$, $\theta_{23}$
and $\delta$).

It is numerically too costly to handle such a large number of
parameters at the same time, hence we analyzed several different
special cases. First, we took a single mass splitting
$\varepsilon_{i}^{2}\neq0$.  If we split two neutrinos with mass $m_i$
into a quasi-degenerate pair of particles with masses
$\sqrt{m_i^2-\varepsilon_i^2/2}$ and $\sqrt{m_i^2+\varepsilon_i^2/2}$
then a new oscillation length $L \propto 1/\varepsilon_i^2$ appears
which is associated to the conversion of active to sterile neutrinos.
Very stringent limits on $\varepsilon_i^2$ in such one parameter
extensions can be derived, of the order of $10^{-11}\textrm{ eV}^{2}$
for $\varepsilon_{1,2}^2$ (from solar neutrino data) and
$10^{-5}\textrm{ eV}^{2}$ for $\varepsilon_{3}^2$ (dominated by
Super-K atmospheric neutrino data).

Next, we considered the case when one mass splitting and one of the
non-standard angles are allowed to take non-zero values at the same
time. As we have shown, in this situation degeneracies of the $\chi^2$
function can occur, implying that from a single experiment in many
cases it will no longer be possible to derive meaningful limits on
individual parameters. These degeneracies can be resolved by
considering data from more than one experiment, accessing
different $P\left(\nu_{\alpha}\rightarrow\nu_{\beta}\right)$.

We then considered the possibility of nullifying the
effects of the $\varepsilon_i$ completely by changing some particular
combinations of the angles $\theta_{ij}$ of our parametrization.
Instead of pursuing the exact form of these rather complex parameter
combinations, we discussed a simpler definition, describing 3 angles
$\varphi_{i}$ associated to rotations between the columns $i$ and
$i+3$ of the quasi-Dirac mixing matrix, such that in the limit where
these angles are equal to $\pi/4$ ($3\pi/4$) the $i+3$($i$) column of
the mixing matrix vanishes and hence the associated neutrino mass
disappears from the oscillation probability formula. We stress that
for these particular parameter combinations no limits on the
$\varepsilon_i$ can be derived from oscillation experiments.  The
regions in the planes $(\varepsilon_i,\varphi_{i})$ which are allowed
by various experiments are shown in figs. (\ref{fig:AtmPhi36}) and
(\ref{fig:ChiSolPhi14}). In this context, it is interesting to note
that the tension between the value of the solar mass scale preferred
by global fits ($\sim7.6\times10^{-3}\textrm{ eV}^{2}$) and the lower
one preferred by solar data ($\sim4\times10^{-3}\textrm{ eV}^{2}$)
might be resolved by a non-zero value for either $\varepsilon_1$ or
$\varepsilon_2$.

Lastly, we considered the possibility that the mass splittings
$\varepsilon_{i}$ are too small to be measured in oscillation
experiments. Even in this scenario, one can have departures from the
lepton-number-conserving Dirac scenario due to the new angles
$\theta_{ij}$ (and phases $\phi_{ij}$). As mentioned above, there is a
large number of such parameters. However, it can be shown that with 3
pairs of neutrinos with the same mass, oscillations will only depend
on a total of 13 combinations of angles and phases. Additionally, if
we focus just on electron and muon neutrinos, this number is further
reduced to 7, corresponding to 6 angles and 1 phase. In the text we
called these parameter combinations $X_{1\cdots 7}$ and stressed that
they can not be identified with $\theta_{12}$, $\theta_{13}$,
$\theta_{23}$ nor $\delta$, as these quantities by themselves are not
physical.  Instead, the 7 $X_i$ correspond to combinations of these
and additional $\theta_{ij}$ angles and $\phi_{ij}$ phases.

In section \eqref{subsect:angs} we made a 7-dimensional scan of these
$X_i$ parameters in the absence of mass splittings. Their exact
definitions, as well as the limits imposed on them by current data can
be found there. Crucially, for Dirac neutrinos there are only 4
parameters. Hence the Dirac limit corresponds to 3 relations among the
7 $X_i$. By testing these relations, we find that
$\min\left(\chi_{\textrm{Dirac}}^{2}\right)-\min\left(\chi^{2}\right)=1.9$,
i.e. current data is compatible with the Dirac scenario. Progress on
tests for quasi-Diracness can be made in the future with a more
precise measurement of $P\left(\nu_{e}\rightarrow\nu_{\mu}\right)$ and
$P\left(\nu_{\mu}\rightarrow\nu_{\mu}\right)$. Thus, more statistics
taken in T2K, MINOS+ or NO$\nu$A and, in particular, the future
precise measurements possible at DUNE should provide more sensitive
probes for this particular setup of quasi-Dirac neutrinos without new
mass scales.

\section*{Acknowledgements}

We would like to thank Mariam Tórtola for patiently explaining to
us, on numerous occasions, several details concerning the various
neutrino oscillation experiments. This work was funded by the Spanish
state through the projects FPA2014-58183-P and SEV-2014-0398 (from
the Ministerio de Economía, Industria y Competitividad), as well as
PROMETEOII/2014/084 (from the Generalitat Valenciana). R.F. was also
financially supported through the grant Juan de la Cierva-formación
FJCI-2014-21651.

\end{document}